\pdfoutput=1
\documentclass[fleqn,usenatbib]{mnras}

\usepackage{newtxtext,newtxmath}

\usepackage[T1]{fontenc}

\usepackage{graphicx}	
\usepackage{gensymb}
\usepackage{hyperref}
\usepackage{booktabs}
\usepackage{supertabular}
\usepackage{caption}
\usepackage{longtable}

\newcommand{\host}{TOI-544}
\newcommand{\planetb}{TOI-544\,b}
\newcommand{\planetc}{TOI-544\,c}
                          
\newcommand{\Msun}{M$_{\odot}$}

\newcommand{\mearth}{M$_{\oplus}$}

\newcommand{\rearth}{R$_{\oplus}$}

\newcommand{\bmass}{2.89 $\pm$ 0.48 }
\newcommand{\brad}{2.018 $\pm$ 0.076 }
\newcommand{\cmass}{21.5 $\pm$ 2.0 }
\newcommand{\bdensity}{$1.93^{+0.30}_{-0.25}$}

\title[TOI-544\,b: a potential water-world inside the radius valley in a two-planet system]{TOI-544\,b: a potential water-world inside the radius valley in a two-planet system}

\author[H. L. M. Osborne et al.]{H. L. M. Osborne,$^{1,2}$\thanks{E-mail: hannah.osborne.19@ucl.ac.uk}
V. Van Eylen,$^{1}$
E. Goffo,$^{3,4}$
D. Gandolfi, $^{3}$
G. Nowak,$^{5,6,7}$
C. M. Persson,$^{8}$ \newauthor
J. Livingston,$^{9,10,11}$ 
A. Weeks,$^{1}$
E. Pall\'e,$^{6,7}$
R. Luque,$^{12}$
C. Hellier,$^{13}$
I. Carleo,$^{6}$ 
S. Redfield,$^{14}$ 
T. Hirano,$^{9,10,11}$ \newauthor
M. Garbaccio Gili,$^{3}$
J. Alarcon,$^{15}$
O. Barrag\'an,$^{16}$
N. Casasayas-Barris,$^{6,7}$ 
M. R. D\'iaz,$^{17}$
M. Esposito,$^{4}$ \newauthor
J. S. Jenkins$^{19,20}$
E. Knudstrup,$^{18}$ 
F. Murgas,$^{6,7}$
J. Orell-Miquel,$^{6,7}$ 
F. Rodler,$^{15}$
L. Serrano,$^{3}$
M. Stangret,$^{6,7}$ \newauthor
S. H. Albrecht,$^{18}$
A. Alqasim,$^{1}$ 
W. D. Cochran,$^{22}$ 
H. J. Deeg,$^{12,23}$
M. Fridlund,$^{8,21}$
A. Hatzes,$^{4}$
J. Korth,$^{24}$ \newauthor
K. W. F. Lam$^{25}$
\\
$^{1}$Mullard Space Science Laboratory, University College London, Holmbury St Mary, Dorking, Surrey RH5 6NT, UK\\
$^{2}$European Southern Observatory, Karl-Schwarzschild-Straße 2, 85748 Garching bei M\"unchen, Germany\\
$^{3}$Dipartimento di Fisica, Universit\'a di Torino, via P. Giuria 1, 10125 Torino, Italy\\
$^{4}$Th\"uringer Landessternwarte Tautenburg, Sternwarte 5, 07778 Tautenburg, Germany\\
$^{5}$Institute of Astronomy, Faculty of Physics, Astronomy and Informatics, Nicolaus Copernicus University, Grudzi\c{a}dzka 5, 87-100 Toru\'n, Poland\\
$^{6}$Instituto de Astrof\'isica de Canarias (IAC), c/ Via Lactea, s/n, 38205 La Laguna, Tenerife, Spain\\
$^{7}$Departamento de Astrof\'isica, Universidad de La Laguna (ULL), 38206 La Laguna, Tenerife, Spain\\
$^{8}$Department of Space, Earth and Environment, Chalmers University of Technology, Onsala Space Observatory, 439 92 Onsala, Sweden\\
$^{9}$Astrobiology Center, 2-21-1 Osawa, Mitaka, Tokyo 181-8588, Japan\\
$^{10}$National Astronomical Observatory of Japan, 2-21-1 Osawa, Mitaka, Tokyo 181-8588, Japan\\
$^{11}$Department of Astronomy, The Graduate University for Advanced Studies (SOKENDAI), 2-21-1 Osawa, Mitaka, Tokyo, Japan\\
$^{12}$Department of Astronomy and Astrophysics, University of Chicago, Chicago, IL 60637, USA\\
$^{13}$Astrophysics Group, Keele University, Staffordshire ST5 5BG, UK\\
$^{14}$Astronomy Department and Van Vleck Observatory, Wesleyan University, Middletown, CT 06459, USA\\
$^{15}$European Southern Observatory, Alonso de Cordova 3107, 7630391 Vitacura, Santiago de Chile, Chile\\
$^{16}$Sub-department of Astrophysics, Department of Physics, University of Oxford, Oxford OX1 3RH, UK\\
$^{17}$Las Campanas Observatory, Carnegie Institution of Washington, Colina el Pino, Casilla 601 La Serena, Chile\\
$^{18}$Stellar Astrophysics Centre, Department of Physics and Astronomy, Aarhus University, Ny Munkegade 120, DK-8000 Aarhus C, Denmark\\
$^{19}$Instituto de Estudios Astrof\'isicos, Facultad de Ingenier\'ia y Ciencias, Universidad Diego Portales, Av. Ej\'ercito 441, Santiago, Chile\\
$^{20}$Centro de Astrof\'isica y Tecnolog\'ias Afines (CATA), Casilla 36-D, Santiago, Chile\\
$^{21}$Leiden Observatory, University of Leiden, PO Box 9513, 2300 RA Leiden, The Netherlands\\
$^{22}$McDonald Observatory and Center for Planetary Systems Habitability, The University of Texas, Austin Texas USA\\
$^{23}$Instituto de Astrofísica de Andalucía (IAA-CSIC), Glorieta de la Astronomía s/n, 18008 Granada, Spain\\
$^{24}$Lund Observatory, Division of Astrophysics, Department of Physics, Lund University, Box 43, 22100 Lund, Sweden\\
$^{25}$Institute of Planetary Research, German Aerospace Center (DLR), Rutherfordstrasse 2, D-12489 Berlin, Germany\\
}

\date{Accepted 2023 December 06. Received 2023 December 01; in original form 2023 May 16}

\pubyear{2023}

\begin{document}
\label{firstpage}
\pagerange{\pageref{firstpage}--\pageref{lastpage}}
\maketitle

\begin{abstract}
We report on the precise radial velocity follow-up of TOI-544 (HD 290498), a bright K star (V=10.8), which hosts a small transiting planet recently discovered by the Transiting Exoplanet Survey Satellite (TESS). We collected 122 high-resolution HARPS and HARPS-N spectra to spectroscopically confirm the transiting planet and measure its mass. The nearly 3-year baseline of our follow-up allowed us to unveil the presence of an additional, non-transiting, longer-period companion planet. We derived a radius and mass for the inner planet, TOI-544\,b, of \brad \rearth\  and \bmass \mearth\  respectively, which gives a bulk density of \bdensity\ g~cm$^{-3}$. TOI-544\,c has a minimum mass of \cmass~\mearth\ and orbital period of 50.1 $\pm$ 0.2 days. The low density of planet-b implies that it has either an Earth-like rocky core with a hydrogen atmosphere, or a composition which harbours a significant fraction of water. The composition interpretation is degenerate depending on the specific choice of planet interior models used. Additionally, TOI-544\,b has an orbital period of 1.55 days and equilibrium temperature of 999\,$\pm$\,14\,K, placing it within the predicted location of the radius valley, where few planets are expected. TOI-544\,b is a top target for future atmospheric observations, for example with JWST, which would enable better constraints of the planet composition.
\end{abstract}




\section{Introduction}

The radius valley describes the region in the size distribution of exoplanets where few planets exist, specifically between 1.5 and 2.0 R$_{\oplus}$. This feature was first observationally identified in \cite{fulton_california-_2017} and \cite{van_eylen_asteroseismic_2018} and a variety of theories have been proposed to explain this gap in planetary radii, including photo-evaporation \citep{owen_evaporation_2017, fulton_california-_2017, van_eylen_asteroseismic_2018} and core-powered mass loss \citep{2018MNRAS.476.2542C,gupta_sculpting_2019,gupta_caught_2021}.
Photo-evaporation models argue that planets generally form with rocky cores and atmospheric layers composed of hydrogen and helium (H-He) of around 1$\%$ by mass. Such planets, with a rocky core and H-He atmosphere, are termed sub-Neptunes and located above the radius valley in radius-period space. In some cases, the intense X-ray flux from the nearby host star strips away these volatile gases, leaving behind a bare rocky core, a so-called super-Earth planet, with a radius placing it below the radius valley \citep{owen_evaporation_2017}.
Core-powered mass loss models are similar in that they predict planets to form with atmospheric layers and be located above the valley and subsequent atmospheric loss reduces the radius and locates the planet below the valley. In the latter model, the energy enabling the mass loss has come from  within the planet itself; stored heat from the formation of the planet escapes from the core and heats the atmospheric layer from the inside, leading to gaseous escape \citep{gupta_sculpting_2019}.
Despite several attempts \citep[e.g.][]{Lopez_2018,owen_evaporation_2017, gupta_sculpting_2019, estrela_evolutionary_2020, gupta_properties_2022,ho_deep_2023} no significant observational evidence has been found that can differentiate between models. There is also an additional complication in the fact that the location of the radius valley seems to change based on other parameters, in particular on stellar mass (see e.g. \cite{petigura_california-kepler_2022}. \cite{Cloutier2020} calculated the occurrence rates of small planets orbiting low mass stars to show that the location of the radius valley shifts to smaller sizes for decreasing stellar mass. They also argue that for planets around lower mass stars the radius valley may have a different formation mechanism or mechanisms, and highlight the need for high-precision RV follow-up of a number of key targets including TOI-544\,b.

Additionally, recent works have suggested that the observed distribution of small planets (particularly those orbiting M-dwarf stars) is the result of a distribution in core composition at formation. Specifically, \cite{luque_density_2022} argue that the small planets around M dwarfs can be separated into super-Earths (with rocky cores) and water worlds (with large fractions of ice or water layers). Other recent detections have also seemed to provide evidence towards this divergence in core compositions. \cite{Piaulet_2022} presents a detailed study of Kepler-138\,d, a small planet with a bulk density which they argue can only be explained with a water world composition (see also e.g. \cite{2022diamondlowe, cadieux_toi-1452_2022}). However, \cite{rogers2023} use models which include atmospheric boil-off shortly after planet formation to show that the group of water-world planets presented in \cite{luque_density_2022} are consistent with atmospheric loss models where a rocky planet with a H-He atmosphere loses its atmospheric layers to become a stripped core.

In this context, we have performed follow-up high-resolution radial velocity observations within the KESPRINT consortium \footnote{\url{https://kesprint.science}} of the small planet TOI-544\,b discovered by the Transiting Exoplanet Survey Satellite \citep[TESS,][]{ricker_transiting_2014}. The planet is particularly interesting since it is located in the middle of the radius valley and has a short period of 1.5 days. It was recently validated by \cite{giacalone2022}, where they present observations from the TESS follow-up programme, who argued that TOI-544\,b is a potentially interesting target for JWST.

In Section~\ref{observations} we present the space-based and ground-based observations of TOI-544, including the extensive RV measurements. In Section~\ref{stellar} we describe our stellar parameter fitting method and results. In Section~\ref{results} we describe both the transit fitting and the RV fitting and in Sections~\ref{composition} to \ref{atmospherics} we explore the composition of the inner planet, its location in relation to the radius valley and potential atmospheric observations of this planet. Finally, we present conclusions in Section~\ref{conclusions}.

\section{Observations}
\label{observations}
\subsection{Space-based Photometry}

\begin{figure*}
    \includegraphics[width=\textwidth]{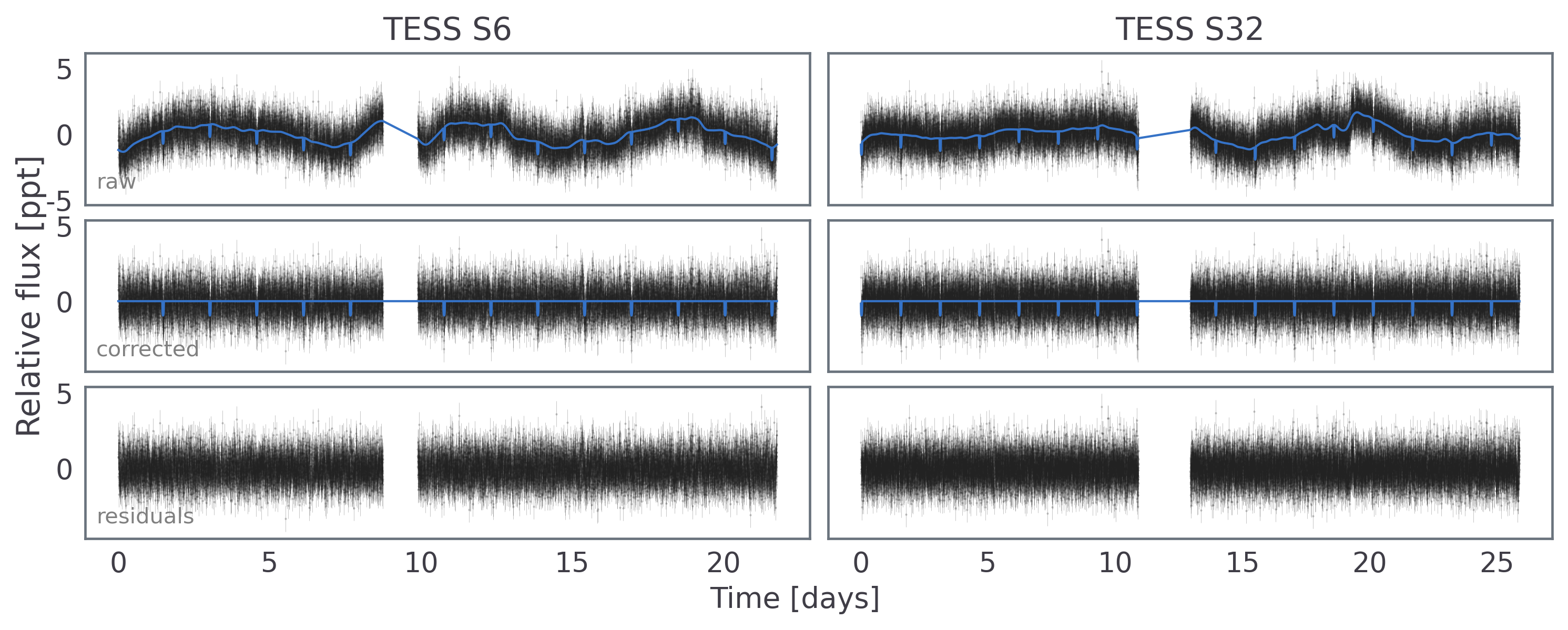}
    \caption{TESS light curves from Sector 6 (left) and 32 (right). SPOC light curves are shown in the upper panel, with a full transit and noise model (see Section~\ref{sec:transitmodel}) shown in blue. In the middle panel, TESS light curves are shown after subtracting the best-fit GP model, and residuals are shown in the bottom row.}
    \label{fig:tess}
\end{figure*}

As part of its all-sky survey, TESS observed the star TIC 50618703 in Sectors 6 and 32. The cadence of observations is 120 s, the time span of Sector 6 (32) is 2018 Dec 12 to 2019 Jan 06 (2020 Nov 19 to 2020 Dec 16), and 44/14691 (817/17977) cadences were omitted due to bad quality flags in Sector 6 (32).
After data reduction though the standard Science Processing Operations Centre \citep[SPOC,][]{jenkins2016,twicken2019} pipeline, likely transits were detected and the planetary candidate was promoted to a TESS object of interest (TOI), named TOI-544, by the TESS team. TESS observations of TOI-544 are shown in Figure \ref{fig:tess}.

\subsection{Ground-based photometry}
\label{WASP_Phot}
\begin{figure}
\includegraphics[width=\columnwidth]{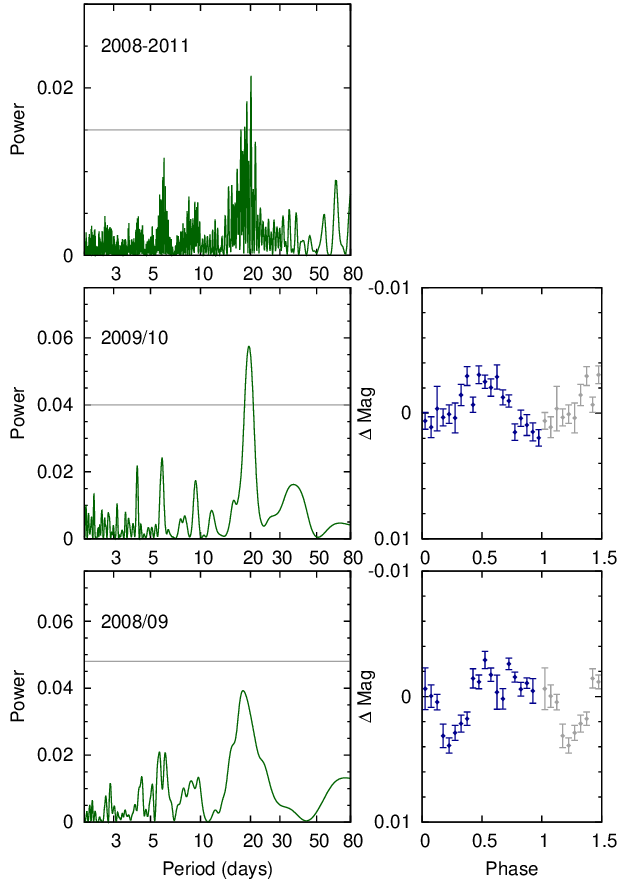}
  \caption{Generalised Lomb-Scargle Periodograms of the WASP-South data for TOI-544. The horizontal line is the estimated 1\%\ false-alarm probability, data after phase 0.5 are in grey. The right-hand panels show the data folded on the 20-d rotational period.}
\label{fig:wasp_toi544}
\end{figure}

Prior to the initial planet detection by TESS, the Wide Angle Search for Planets (WASP) survey \citep{2006PASP..118.1407P} observed the field of TOI-544 between 2008 and 2011, obtaining 18\,000 photometric datapoints using Canon 200-mm, f/1.8 lenses with a 400--700 nm filter and CCDs with a plate scale of $13.7\arcsec$/pixel. The data have a typical 15-min cadence and covered an observing season of $\sim$\,100 nights in each year. TOI-544 is by far the brightest star in the 48-arcsec extraction aperture, with a V magnitude of 10.777$\pm$0.017. We make use of this archival data by searching the resulting WASP lightcurves for a stellar rotational modulation, adopting the methods described in \citet{2011PASP..123..547M}. We find a significant modulation at a period of 20 $\pm$ 1 days and an amplitude of up to 3 mmag. In the combined dataset the modulation is significant to a level of $>$99\%\ confidence. Figure \ref{fig:wasp_toi544} shows Generalised Lomb-Scargle periodograms, adapted as described in \cite{Maxted_2011}, of the WASP data for TOI-544. The 1$\%$ false-alarm probability is shown by the horizontal line and the panels to the right show the data folded on the 20 day rotation period. 

\subsection{High-resolution HARPS and HARPS-N spectroscopy}

We obtained a total of 108 high-resolution ($R\,=\,\lambda/\Delta\lambda\,\approx\,115\,000$) spectroscopic observations of TOI-544 using the High Accuracy Radial velocity Planet Searcher (HARPS) mounted on the European Southern Observatory (ESO) 3.6-m telescope at La Silla Observatory. One hundred and six spectra were acquired between December 2020 and March 2022 as part of our large observing programme 106.21TJ.001 (PI: Gandolfi).
There are an additional 2 observations taken during April 2019 as part of observing programme 0103.C-0442(A) (PI: Diaz).
Each observation has an exposure time between 1500\,s and 2700\,s, leading to a median signal-to-noise (S/N) ratio of 53.7 per pixel at 550~nm (Table~\ref{table-TOI-0544-3p6_harps-0108-drs-complete_output}), and a median RV uncertainty of $\sim$1.6\,m\,s$^{-1}$, as extracted using the SpEctrum Radial Velocity Analyser {\tt SERVAL} (see below; Table~\ref{table-TOI-0544-3p6_harps-0108-srv-complete_output}).

TOI-544 was also observed using the High Accuracy Radial velocity Planet Searcher for the Northern hemisphere \citep[HARPS-N;][]{2012SPIE.8446E..1VC} mounted at the 3.58-m Telescopio Nazionale Galileo (TNG) of Roque de los Muchachos Observatory in La Palma, Spain, between April 2019 and December 2020, resulting in a total of 14 high-resolution ($R\,\approx\,115\,000$) spectra from programmes CAT19A\_162
and CAT19A\_97 (PIs: Nowak and Casasayas-Barris). Each observation has an exposure time between 1230\,s and 2400\,s, a median S/N ratio of 52 per pixel at 550~nm (Table~\ref{table-TOI-0544-tng_harpn-0014-drs-complete_output}), and a median RV precision of $\sim$1.2\,m\,s$^{-1}$, as extracted using {\tt SERVAL} (see below; Table~\ref{table-TOI-0544-tng_harpn-0014-srv-complete_output}). We include these 14 observations in our RV analysis. The total number of HARPS and HARPS-N spectra is thus 122.

Versions 3.8 and 3.7 of the Data Reduction Software \citep[{\tt DRS}; ][]{2002A&A...388..632P,2007A&A...468.1115L} were used to reduce the HARPS and HARPS-N spectra, respectively, and extract absolute RVs by cross-correlating the spectra with a K5 numerical mask \citep{1996A&AS..119..373B}, along with three diagnostics of the cross-correlation function (CCF), namely, the full width at half maximum (FWHM), the bisector inverse slope (BIS), and the contrast. We also used the {\tt SERVAL} code \citep{2018A&A...609A..12Z} to measure relative RVs and extract two additional activity diagnostics, namely, the differential line width (dLW) and the chromatic index (CRX). {\tt SERVAL} implements a template-matching algorithm that is suitable to derive precise Doppler measurements for M- and late K-type stars when compared to the CCF technique employed by the {\tt DRS} \citep[see, e.g.,][]{2021A&A...645A..41L,2022NatAs...6..736S,2023ApJ...955L...3G}. 

We finally extracted the H$\alpha$, Na~D1 \& D2, and Ca\,{\sc ii} S-index activity indicators using the Template Enhanced Radial velocity Re-analysis Application \citep[TERRA;][]{2012ApJS..200...15A}. The standard deviation of the RV data is $\sim$7.5\,m\,s$^{-1}$ for HARPS and HARPS-N and in both cases of using the {\tt SERVAL} and {\tt DRS} codes. The absolute ({\tt DRS}) and relative ({\tt SERVAL}) RV measurements are given in Appendix~\ref{Appendix}, along with the stellar activity indicators and line profile diagnostics. For the analysis presented in Sects.~\ref{GLS} and \ref{RV_Analysis} we used the {\tt SERVAL} RV measurements due to the lower jitter terms and root-mean-square of the fit residuals with respect to the {\tt DRS} RVs.

\section{Stellar Parameters}
\label{stellar}

\begin{table*}
\caption{Stellar parameters for TOI-544, modelled with {\tt {SpecMatch-Emp}} and {{\tt {SME}}, and {\tt {BASTA}} and {\tt {astroARIADNE}.}} }   
\label{Table:stellarparameters}
\begin{tabular}{llllll}
    \hline
    Stellar Parameters & & & & &  \\ 
    \hline
    Identifiers  & TOI-544 & & & & \\
    & TIC 50618703 & & & & \\
    & HD\,290498 & & & & \\
    Coordinates$^1$ & 05:29:09.62 -00:20:34.43 &&&&\\
    Magnitudes$^1$ & V = 10.777 $\pm$ 0.017 &&&& \\
    &TESS = 9.6504 $\pm$ 0.006  &&&&\\ 
    
    \hline
    Spectroscopic Parameters &&&&& \\
    \hline
    Method  & $T_\mathrm{eff}$  & $\log g_\star$ & [Fe/H]   &   $V \sin i$ & Luminosity   \\  
    & (K)  &(cgs) &(dex)       & (km~s$^{-1}$)  & (L$_{\odot}$) \\
    \hline
    {\tt {SpecMatch-Emp}}$^a$   & $4169 \pm 70$  & $4.68\pm 0.12$ & $-0.17\pm0.09$   &\ldots  & \ldots\\
    {\tt {SME}}  &  $4248 \pm 130$  & $4.60\pm 0.12$ & $-0.15\pm0.08$   &   $2.3 \pm 0.8$ &  0.13 $\pm$ 0.02 \\
    \hline 
    & RA  & DEC & Parallax   &   G &    \\  
    \hline
    {\tt {GAIA$^2$}}  & 82.2900716  & -0.3428983 & 24.44 $\pm$ 0.02   &   10.41430 $\pm$ 0.00059 &  \\
    \hline 
    Derived Parameters &&&&& \\
    \hline
    Method    & $M_\star$  &     $R_\star$  & Distance & Age    & \\
    & ($M_{\odot}$)  & ($R_{\odot}$)  & (Pc) & (Gyear)  &\\
    \hline
    {\tt {BASTA}}$^b$      &   $0.630\pm 0.018$  &  $0.624\pm0.013$ & 40  $\pm$ 1  & 9  $\pm7$ &\\
    {\tt {MIST}}      &   $0.645\pm 0.021$  &  $0.617\pm0.015$ & 40.9  $\pm$ 0.03  & $5^{+5.3}_{-3.1}$ &\\
    {\tt {astroARIADNE}}   &  $0.651^{+0.015}_{-0.026}$  &  $0.630^{+0.044}_{-0.017}$    & \ldots & \ldots &   \\
    \hline 
    \end{tabular} \\
\textit{a}{ Adopted as priors for the stellar mass and radius modelling.}
\textit{b}{ Adopted as our final stellar parameter values to be used in analysis of the planetary system.\textit{1} Taken from TESS Input Catalogue (TIC) version 8 \citep{TIC2019}. \textit{2} Taken from Gaia DR3 \citep{DR3}.}
\end{table*}

Stellar parameters for TOI-544 were calculated using BASTA \citep{BASTA} run on the co-added HARPS spectra, as follows. First, Gaia magnitude $G$, RA, DEC, and Parallax $\varpi$ were taken from Gaia DR3 \citep{DR3}.  
T$_{eff}$ and [Fe/H] were determined using the empirical software SpecMatch-Emp \citep{specmatch2017} which compares the observed data to a library of FGKM stars. 
We also compared the Spec-Match-Emp results with SME\footnote{\url{http://www.stsci.edu/~valenti/sme.html}} \citep[Spectroscopy Made Easy;][]{vp96, pv2017}. SME computes synthetic spectra with line data from VALD\footnote{\url{http://vald.astro.uu.se}}  \citep{Ryabchikova2015} and a chosen stellar atmosphere grid, in our case  Atlas12 \citep{Kurucz2013}, which is fitted to the observed spectra. 
The macro- and micro-turbulent velocities, $V_{\rm mac}$ and $V_{\rm mic}$, were held fixed to 1.5~km~s$^{-1}$ and 0.5~km~s$^{-1}$, respectively \citep{gray08}. 
A more detailed description   of the modeling procedure can be found in \citet{2017A&A...604A..16F} and \citet{2018A&A...618A..33P}.  The results were in very good agreement 
with SpecMatch-Emp within $1~\sigma$. Table \ref{Table:stellarparameters} gives the spectroscopic  parameters for TOI-544 modelled with {\tt {SME}} and {\tt {SpecMatch-Emp}}, we select the parameters using ({\tt {SpecMatch-Emp}}) for our modelling of stellar mass and radius.

An age and metallicity independent prior was placed on stellar mass following the standard Salpeter Initial Mass Function, and reddening and dust were accounted for using the \verb+‘Bayestar’+ dustmap \citep{bayestar2019}. A prior was also set for [Fe/H] allowing for the parameter space to be searched only within an absolute tolerance range of 0.5 dex  of the input value. For the isochrones, we used the latest version of BASTI (BAg of STellar Isochrones) \citep{basti1}, set to the `Diffusion’ science case, described in \citet{basti2}, to account for diffusion processes in low-mass stars. 

BASTA determines model dependent parameters using a Bayesian approach, detailed in \citet{SA2015} and following the formalism of \citet{serenelli2013}. 

We also used the Python {\tt isochrones} \citep{2015ascl.soft03010M} interface to the MIST stellar evolution models \citep{2016ApJ...823..102C} using the same inputs as for BASTA, which resulted in very good agreement for the stellar mass and radius (within $\sim$0.5$\sigma$).

We compared our results from BASTA with the software astroARIADNE\footnote{\url{https://github.com/jvines/astroARIADNE}}
\citep[][]{2022arXiv220403769V}. This python code fits the observed spectral energy distribution via broad band photometry to
atmospheric model grids to obtain the stellar radius. We fitted the    bandpasses 
$G G_{\rm BP} G_{\rm RP}$   (Gaia eDR3),   {\it WISE} W1-W2,     
$JHK_S$ magnitudes ({\it 2MASS}),  the Johnson $B$ and $V$ magnitudes (APASS), and 
the Gaia eDR3 parallax. We used the    {\tt {Phoenix~v2}} 
\citep{2013A&A...553A...6H} and the {\tt {BtSettl}} \citep{2012RSPTA.370.2765A} atmospheric models.  The final radius was computed
with Bayesian Model Averaging and the errors with a sampling method for conservative 
uncertainties as described in \citet[][]{2022arXiv220403769V}. In this way we
obtained a stellar radius of $0.630^{+0.044}_{-0.017}$~$R_{\odot}$. The stellar mass was
computed with ARIADNE and the MIST \citep{2016ApJ...823..102C} isochrones and was
found to be $0.651^{+0.015}_{-0.026}$~$M_{\odot}$. 
 Table \ref{Table:stellarparameters} gives the stellar parameters derived from our analysis, as well as the comparisons with ARIADNE, which are within 1 $\sigma$ of our results from BASTA.

\section{Analysis and Results}
\label{results}
\subsection{Transit model}
\label{sec:transitmodel}

We jointly fit the SPOC Pre-Search Data Conditioning Simple Aperture Photometry (PDCSAP) light curves from TESS Sectors 6 and 32 using the \texttt{PyMC3} \citep{pymc3}, \texttt{exoplanet}\footnote{\url{https://docs.exoplanet.codes/en/stable/}} \citep{exoplanet}, \texttt{starry} \citep{luger18}, \texttt{celerite2} \citep{celerite1,celerite2} software packages. 

To account for stellar activity signals and instrumental Systemics we included a Gaussian Process \citep[GP,][]{RasmussenWilliams2005} model, using a Mat\'ern-3/2 covariance function. We placed Gaussian priors on the stellar mass and radius based on the results in Table \ref{Table:stellarparameters}. We also placed Gaussian priors on the limb darkening coefficients based on interpolation of the parameters tabulated by \citet{Claret2012} and \citet{Claret2017}, propagating the uncertainties in the stellar parameters in Table \ref{Table:stellarparameters} via Monte Carlo simulation. 

Visual inspection of the TESS lightcurves reveals quasiperiodic variability with a $\sim$1 ppt amplitude, and their Lomb-Scargle periodograms reveal peaks at $\sim$8 and $\sim$10 days, which are likely the first harmonic of the stellar rotation signal modulo instrumental noise. We thus placed loose Gaussian priors on the GP amplitude and timescale hyperparameters of 1.0\,$\pm$\,0.5 ppt and 10\,$\pm$\,5 days, respectively. 

We used separate white noise parameters for each TESS sector to account for the possibility of differences in photometric precision, which could potentially arise from different background light conditions or different phases of the spacecraft's operational lifetime. We used the gradient-based {\tt BFGS} algorithm \citep{NoceWrig06} implemented in {\tt scipy.optimize} to find initial maximum a posteriori (MAP) parameter estimates. We used these estimates to initialize an exploration of parameter space via ``no U-turn sampling'' \citep[NUTS,][]{HoffmanGelman2014}, an efficient gradient-based Hamiltonian Monte Carlo (HMC) sampler implemented in {\tt PyMC3}. 

We sampled four chains with 4500 tuning iterations and 3000 additional draws, for a total of 12,000 samples after burn-in; the resulting chains were well-mixed according to a Gelman-Rubin statistic \citep{GelmanRubin1992} value of $<$1.01, and the sampling error was $\lesssim$1\%, suggesting a sufficient number of independent samples had been collected. 

\begin{figure}
    \includegraphics[width=\columnwidth]{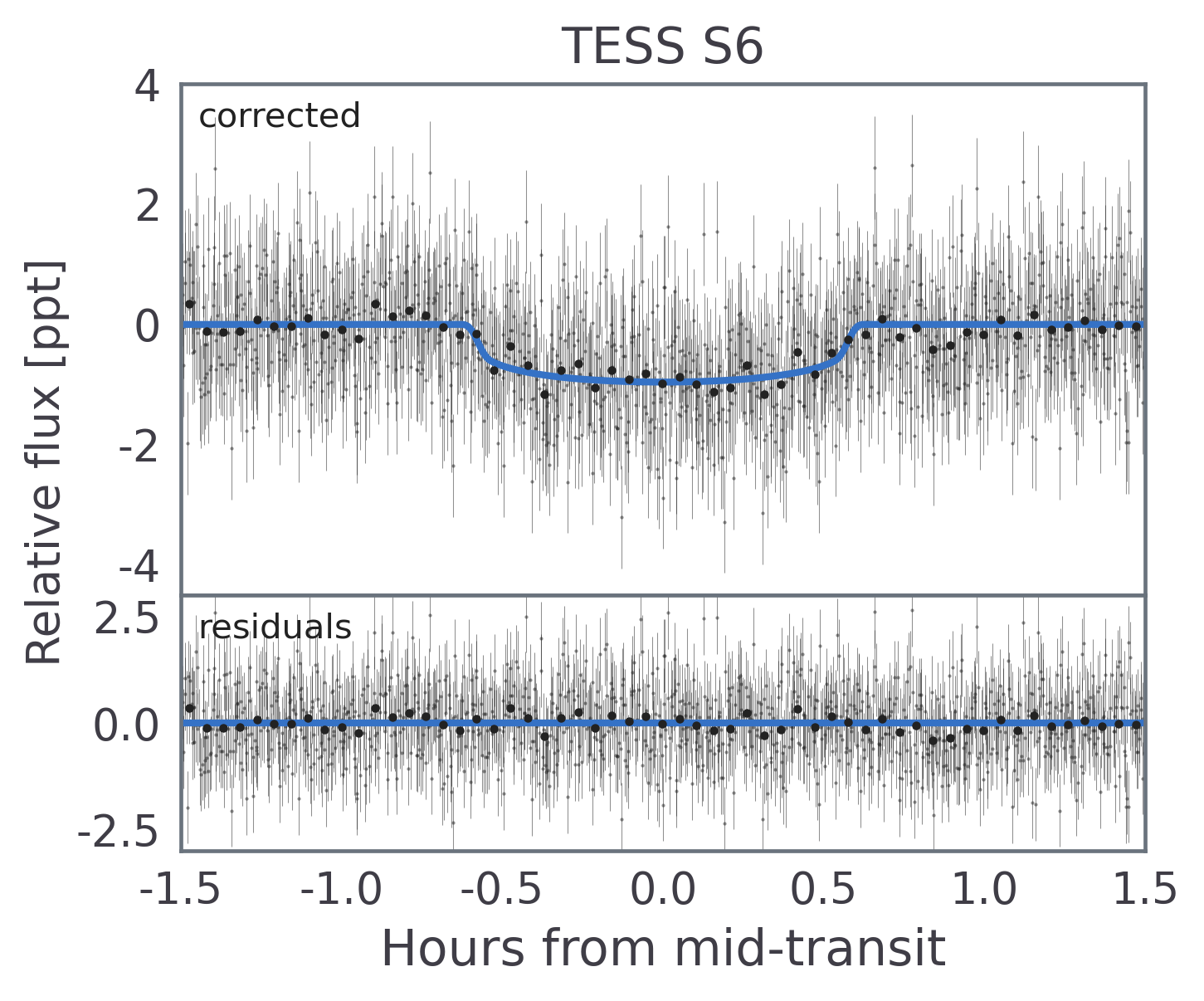}
    \includegraphics[width=\columnwidth]{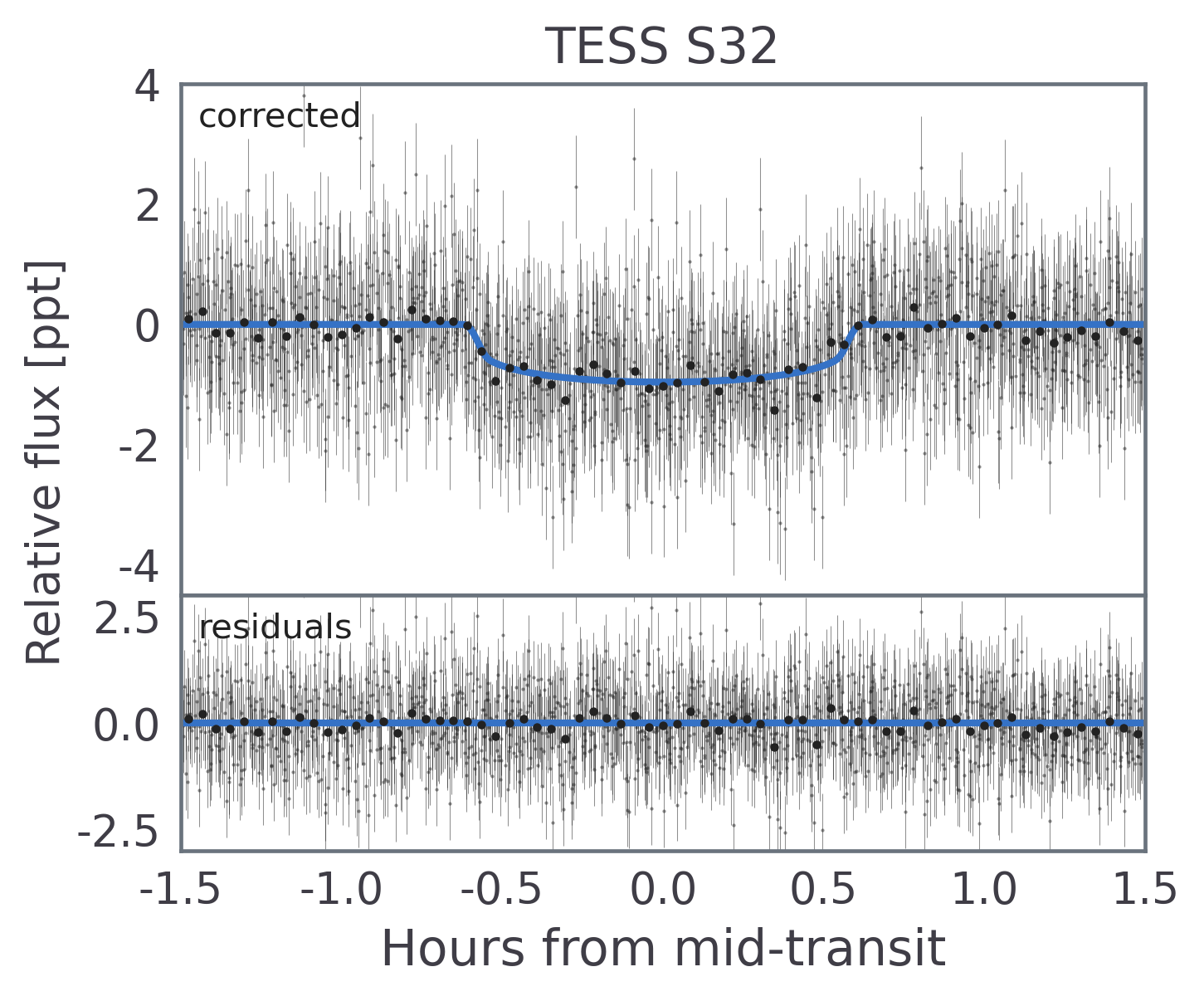}
    \caption{Phase-folded TESS photometry from Sector 6 (top) and Sector 32 (bottom), with the best-fit transit model in blue, the lower panel of each shows the residuals to the best fit model. The larger black points show the data binned by a factor of 30.}
    \label{fig:tess-folded}
\end{figure}

The phase-folded TESS photometry from Sector 6 and 32, along with the best-fit transit model, is shown in Figure \ref{fig:tess-folded} and the results of the transit fit are given in Table \ref{tab:allresults}.

\subsection{Frequency analysis of HARPS data}
\label{GLS}

\begin{figure*}
\centering
\includegraphics[width=0.95\linewidth]{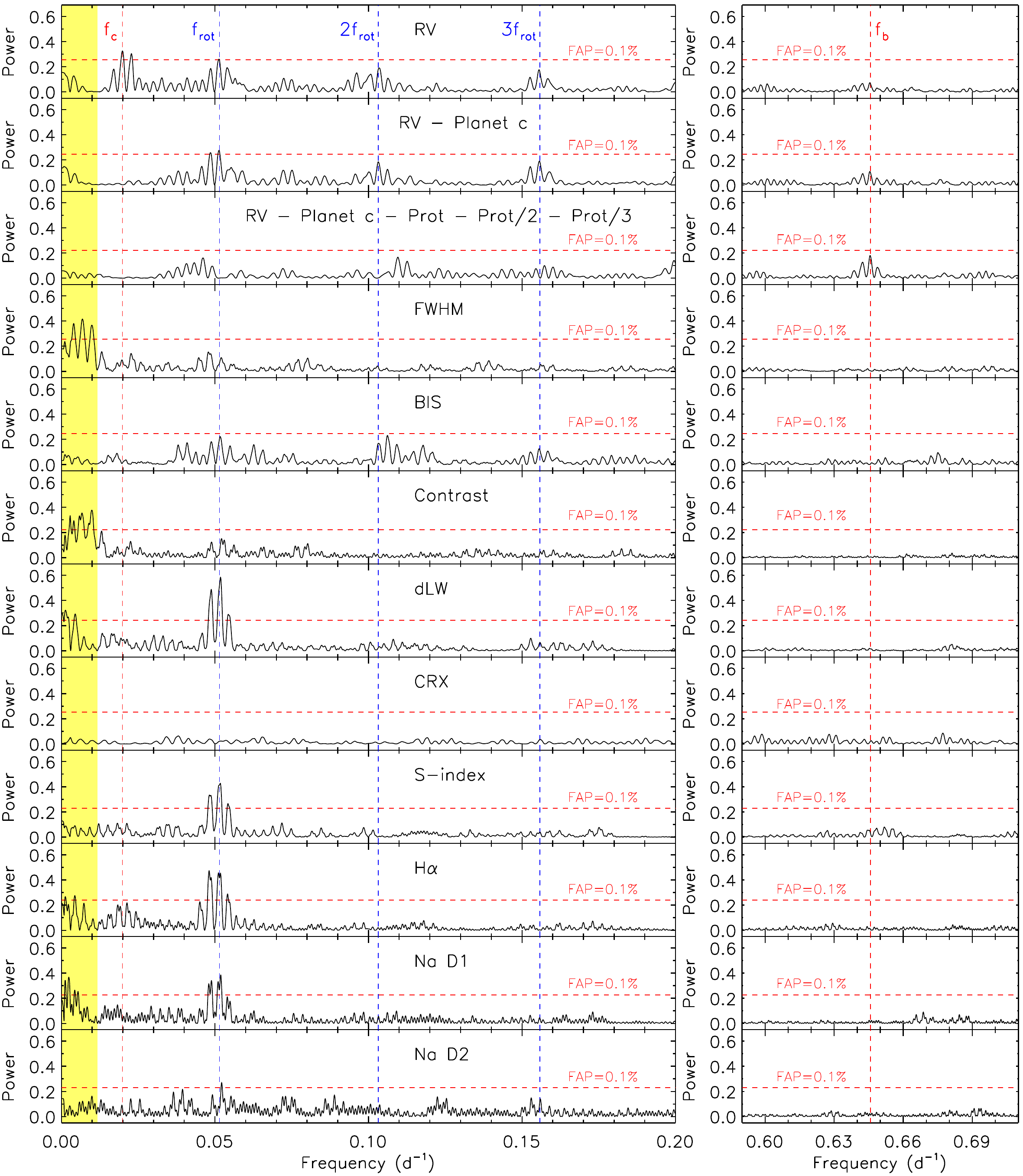}
    \caption{Generalised Lomb-Scargle periodograms of the {\tt SERVAL} RV measurements and activity indicators of \host. The right and left columns cover two frequency ranges encompassing the orbital frequency of \planetc, the stellar rotation frequency and its first two harmonics (left panels), and the orbital frequency of \planetb\ (right panels). The Doppler signals from the two planets are marked as vertical red dashed lines at f$_\mathrm{c}$\,$\approx$\,0.02\,d$^{-1}$ (\planetc) and f$_\mathrm{b}$\,$\approx$\,0.646\,d$^{-1}$ (\planetb). Vertical, blue dashed lines indicate the stellar rotation frequency (f$_\mathrm{rot}$\,$\approx$\,0.051\,d$^{-1}$) and its first two harmonics (2\,f$_\mathrm{rot}$\,$\approx$\,0.102\,d$^{-1}$ and 3\,f$_\mathrm{rot}$\,$\approx$\,0.153\,d$^{-1}$). The horizontal red dashed line mark the 0.1\% false alarm probability. The shaded yellow band highlights the frequency range encompassing the long-period activity signals that we significantly detected in the the FWHM, contrast, dLW, H$\alpha$, and Na~D1, and that are likely related to spot evolution, along with their 1-year aliases (Sect.~\ref{GLS}), last paragraph. \emph{From top to bottom:} the {\tt SERVAL} RV measurements; RV residuals following the subtraction of the Doppler reflex motion induced by \planetc; RV residuals after subtracting the signals of the star rotation, its first two harmonics, and \planetc; FWHM; BIS; contrast; dLW; CRX; S-index; H$\alpha$; Na~D1; Na~D2.  
    \label{figure-TOI-0544-gls_periodograms}}
\end{figure*}

We performed a frequency analysis of our RV time-series to search for the Doppler reflex motion induced by the transiting planet discovered by TESS, spectroscopically confirm its planetary nature, and possibly unveil the presence of additional signals induced by stellar activity and/or additional orbiting companions. In order to avoid having to account for the RV offset between the two spectrographs, we did not include the 14 HARPS-N Doppler data points, and used only the 108 HARPS measurements, which cover a baseline of $\sim$1051\,d (nearly 3 years), implying a frequency resolution of $\sim$0.00095\,d$^{-1}$.

Figure~\ref{figure-TOI-0544-gls_periodograms} displays the generalised Lomb-Scargle \citep[GLS;][]{2009A&A...496..577Z} periodograms of the HARPS {\tt SERVAL} RV measurements and of the activity indicators in two frequency ranges. The panels to the left show the periodograms between 0.0 and 0.2\,d$^{-1}$, a range that includes the frequencies at which we expect to detect the Doppler signals induced by active regions corotating with the star. The panels to the right display the periodograms in the range 0.59-0.71\,d$^{-1}$, which encompasses the transit frequency of TOI-544\,b. We note that the gap between these frequencies shows nothing of note and is excluded. For each panel, the horizontal dashed line mark the false-alarm probability (FAP) of 0.1,\%, as derived using the bootstrap method \citep{Murdoch1993,Kuerster1997}. We considered a peak to be significant if its FAP\,$<$\,0.1\%. 

The GLS periodogram of the HARPS RVs (Fig.~\ref{figure-TOI-0544-gls_periodograms}, upper left panel) displays its most significant peak at $\sim$0.02\,d$^{-1}$ ($\sim$50\,d). This peak does not appear in any other periodograms of the activity indicators\footnote{We note that there are two peaks in the periodogram for H$\alpha$ which are symmetrically located around the peak at $\sim$0.02\,d$^{-1}$. However, neither these are at the same frequency as the signal, nor they are significant (FAP\,$>$\,0.1\,\%)}, providing evidence that the 50-d signal is very likely caused by an additional companion, which we refer to as TOI-544\,c throughout the paper.

We removed the Doppler reflex motion induced by TOI-544\,c fixing period and phase to the values derived from the periodogram analysis, while fitting for the RV semi-amplitude and systemic velocity. The periodogram of the RV residuals following the subtraction of the Doppler reflex motion induced by TOI-544\,c (Fig~\ref{figure-TOI-0544-gls_periodograms}, second left panel) shows 3 equally spaced peaks at $\sim$0.051\,d$^{-1}$ ($\sim$19.4\,d), $\sim$0.102\,d$^{-1}$ ($\sim$9.8\,d), and $\sim$0.153\,d$^{-1}$ ($\sim$6.5\,d). The former is significantly detected also in the periodograms of the dLW, CCF-BIS, S-index, H$\alpha$, Na D1, and Na D2 lines (Fig~\ref{figure-TOI-0544-gls_periodograms}, lower left panels), implying that this is very likely due to stellar activity. We interpreted the peak at $\sim$19.4\,d as the stellar rotation period, in excellent agreement with the WASP results (Sect.~\ref{WASP_Phot}). The two peaks at $\sim$9.8\,d and $\sim$6.5\,d are the first and second harmonics of the stellar rotation period, which are likely caused by the presence of spots equally spaced in longitude and/or by the non-coherent nature of the activity-induced signal.

We removed the stellar signal by fitting 3 Fourier components at the stellar rotation period and its first two harmonics. The GLS periodogram of the RV residuals (Fig.~\ref{figure-TOI-0544-gls_periodograms}, third right panel) shows its strongest power at $\sim$0.646\,d$^{-1}$ ($\sim$1.55\,d), the orbital frequency $f_\mathrm{b}$ of the transiting planet TOI-544\,b. Although the FAP of this feature anywhere in the frequency range of the periodogram is higher than 0.1\,\%, the presence of a peak at a known frequency, i.e., the transit frequency, provides strong evidence that this signal is due to planet~b. We estimated the FAP at the orbital frequency of TOI-544\,b using the windowing bootstrap method described in \citet{2019dmde.book.....H}. Briefly, we estimated the bootstrap FAP over a $\Delta \nu$\,=\,0.1\,d$^{-1}$ wide frequency window centered on $f_\mathrm{b}$. We successively narrowed the spectral window at steps of 0.01\,d$^{-1}$ for 10 additional bootstrap randomizations, down to $\Delta \nu$\,=\,0.01\,d$^{-1}$. The fit of the FAP versus window size, extrapolated to the intercept (i.e., the zero window length), yields a FAP\,=\,0.004\,\% at $f_\mathrm{b}$, spectroscopically confirming the planetary nature of the transit signal discovered by TESS.

The periodograms of the FWHM, contrast, dLW, H$\alpha$, and Na D1 (Fig.~\ref{figure-TOI-0544-gls_periodograms}, yellow strip) display significant peaks at frequencies $\lesssim$\,0.01 d$^{-1}$ ($\gtrsim$\,100\,d), which are equally spaced by about 1/365\,$\approx$\,0.0027~d$^{-1}$, i.e., the seasonal sampling of our time series. Hence, most of these peaks are 1-year aliases of true signals with periods of about 100-250\,d. This range includes the evolution timescale of active regions $\mathrm{\lambda_e}\,=\,112^{+28}_{-29}$~d, as inferred by our multi-dimensional Gaussian process analysis (see Sect.~\ref{RV_Analysis}), suggesting that these signals are associated to long-term stellar variability and spot evolution.

\subsection{Radial velocity analysis}
\label{RV_Analysis}

We modelled the RV data using the code \texttt{pyaneti} \citep{pyanetii,pyanetiii}, which implements a multi-dimensional GP to help account for the impact of stellar activity. The implementation of GPs for this purpose is described in detail in \cite{rajpaul_gaussian_2015}. Essentially, this method models the RVs and an activity indicator of choice, assuming that the same GP, $\mathrm{G(t_i,t_j)}$, can describe both of them. We used the S-index as the activity indicator to model alongside the RVs. As the RVs and the S-index of TOI-544 show a significant stellar rotation signal (Fig.~\ref{figure-TOI-0544-gls_periodograms}), we chose the quasi-periodic (QP) kernel~for~the~GP

\begin{equation}\label{eq:qpkernel}
G \left(t_i, t_j \right) =  A^2 exp \left[ \frac{\sin^2{\left[\pi \left(t_i - t_j\right) / P_{GP}\right]}}{2{\lambda_p}^2} - \frac{ \left(t_i - t_j\right)^2}{{2\lambda_e}^2} \right],
\end{equation}

where $\mathrm{P_{GP}}$ is the GP characteristic period (which here represents the stellar rotation period), $\mathrm{\lambda_p}$ is the inverse of the harmonic complexity (related to distribution of active regions on the stellar surface), and $\mathrm{\lambda_e}$ is the long-term evolution timescale (the lifetime of active regions on the stellar disk). The two-dimensional GP used to model the system is then given by
\begin{equation}
RV = A_0 G + A_1 dG 
\end{equation}
and
\begin{equation}
S\text{-}\mathrm{index} = A_2 G + A_3 dG
\end{equation}
where $A_0$, $A_1$, and $A_2$ are GP hyperparameters in the form of amplitudes that work as a scale factor that determines the typical deviation from the mean function, and $dG$ is the time derivative of our GP function, $\mathrm{G(t_i,t_j)}$. From first principles, $A_3 \equiv 0$, see \cite{rajpaul_gaussian_2015} for more details. We used the stellar parameters listed in Table~\ref{Table:stellarparameters}, and informative Gaussian priors on the orbital period and time of mid-transit for planet b based on those found in the transit fit (Sect.~\ref{sec:transitmodel}). We used the orbital period and time of inferior conjunction we derived from our GLS periodogram to place uniform priors for planet c (see Fig.~\ref{figure-TOI-0544-gls_periodograms} and Sect.~\ref{GLS} for description of this search). We adopted a uniform prior on $P_{GP}$ centered on the stellar rotation period found by the WASP photometry, and wide uniform priors on the remaining model parameters. \texttt{Pyaneti} infers the systemic velocity (aka offset) for each instrument, the exact values found by \texttt{Pyaneti} are given in Table \ref{tab:allresults}. We then performed a Markov chain Monte Carlo (MCMC) analysis, fitting for an eccentric orbit for both planets. We sampled the parameter space with 500 Markov chains and used the final set of 5000 steps with a thin factor of 10 to produce our posterior distributions, which led to a total of 250000 independent points for each sampled parameter.

We find that TOI-544\,b has an orbital period of $\sim$1.55 days and a $K_\mathrm{b}$ amplitude (the radial velocity semi-amplitude) of 2.66\,$\pm$\,0.44~m\,s$^{-1}$, equating to a planet mass of \bmass\,M$_\oplus$. The eccentricity of TOI-544\,b is found to be $0.35^{+0.14}_{-0.12}$ and its argument of periastron is $41.2^{+17}_{-26}$~degrees. TOI-544\,c has a period of 50.1\,$\pm$\,0.2\,days, $K_c$ amplitude of 5.21\,$\pm$\,0.57 m\,s$^{-1}$, implying a minimum mass of \cmass\,M$_\oplus$. The eccentricity of TOI-544\,c is found to be 0.30\,$\pm$\,0.09 and argument of periastron is $17^{+21}_{-18}$ degrees.  

We assessed the significance of the eccentric solutions for the two planets by creating 5000 sets of synthetic RV time-series that sample the best-fitting circular solutions at the epochs of our real observations. We added Gaussian noise at the same level of our RV measurement uncertainties and fitted the simulated time series allowing for non-zero eccentricities. For \planetb\ there is a $\sim$3.5\% that a best-fitting eccentric solution with e $\ge$ 0.35 could arise by chance if the orbit is actually circular. As for \planetc, the probability that noise can account for e\,$\ge$\,0.30 is only 0.2\,\%. Assuming a significance level of 1\%, the eccentric solution for \planetb\ is likely not real, while the eccentricity of \planetc\ is real. Therefore, our adopted results are for the case where the inner planet is on a circular orbit and outer planet is on an eccentric orbit, i.e., TOI-544\,b has an orbital period of $\sim$1.55~days and a $K_\mathrm{b}$ amplitude of 2.17\,$\pm$\,0.36~m\,s$^{-1}$, equating to a planet mass of $M_\mathrm{b} =$ \bmass\,M$_\oplus$,  TOI-544\,c has a period of 50.1\,$\pm$\,0.2~days, $K_\mathrm{c}$ amplitude of 5.36\,$\pm$\,0.55~m\,s$^{-1}$, implying a minimum mass of $M_\mathrm{c} \sin i_\mathrm{c}$ = \cmass\,M$_\oplus$. The eccentricity of TOI-544\,c is found to be $e_\mathrm{c}$\,=\,0.32\,$\pm$\,0.09 and argument of periastron is $\omega_\mathrm{c}\,=\,12^{+19}_{-17}$ degrees. 

\begin{figure*}
	\includegraphics[width=2\columnwidth]{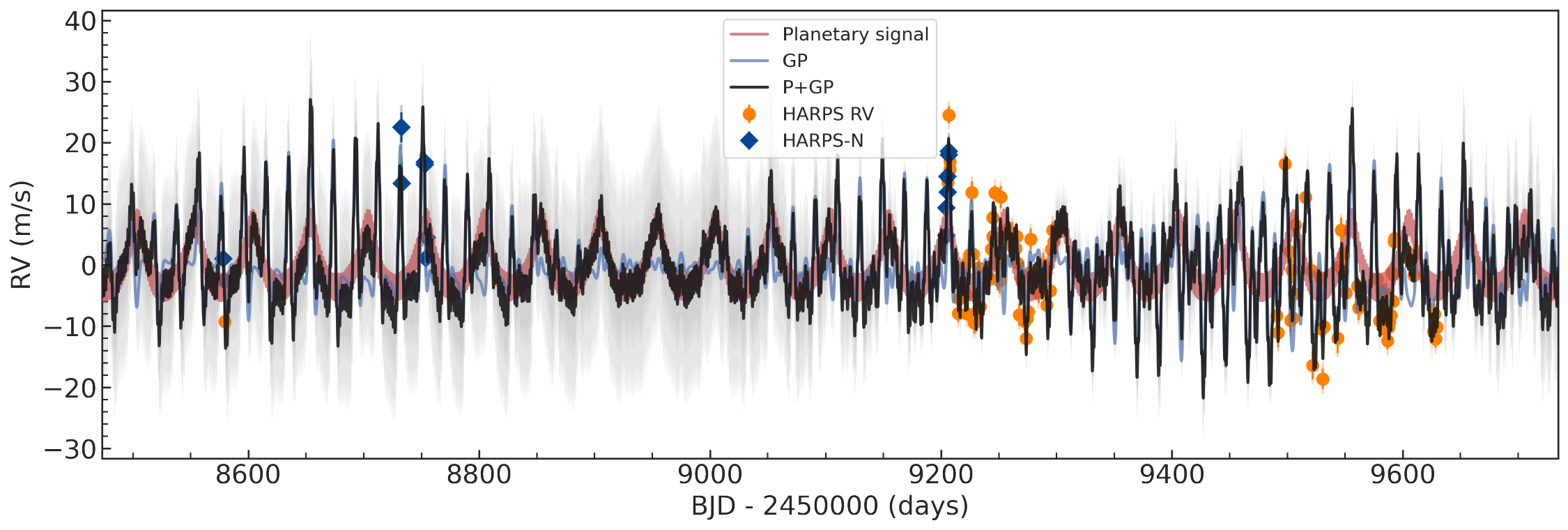}
    \caption{Best-fit 2-planet Keplerian orbital model for TOI-544. HARPS data shown by orange circles, and HARPS-N data shown by blue diamonds, both shown as a function of time. The best-fit model for the planet signals is shown in red, the GP model in blue, and the combined planets and GP shown in black. The dark and light shaded areas showing the 1$\sigma$ and 2$\sigma$ credible intervals of the corresponding GP model, respectively}.
    \label{fig:toi544_timeseries}
\end{figure*}

\begin{figure*}
	\includegraphics[width=2\columnwidth]{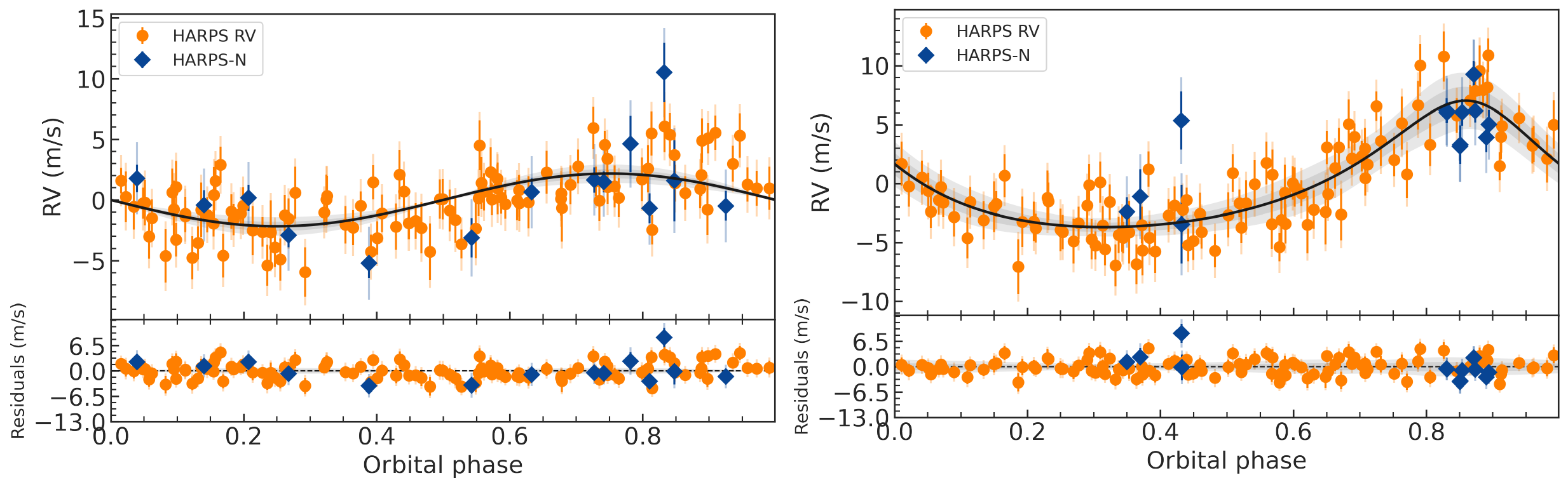}
    \caption{The phase-folded RV data from HARPS (orange circles) and HARPS-N (blue diamonds) alongside the best-fit planet model (with the effect of the other planet and the GP model subtracted from each). TOI-544\,b is on the left and TOI-544\,c on the right. The lower part shows the residuals from the fit. There appears to be no trends visible in the residuals, and the augmented sinusoid representative of an eccentric orbit is present in both phase-folded plots}.
    \label{fig:toi544_phasefolded}
\end{figure*}

The priors used for all parameters are show in Table \ref{tab:priors}, and the results, showing the median value and 68 $\%$ credible interval for each parameter, are given in Table \ref{tab:allresults}. The best fit model alongside the data are shown, as a function of time, in Figure \ref{fig:toi544_timeseries} and phase-folded for each planet in Figure \ref{fig:toi544_phasefolded}. 

We additionally chose to model the RV data using several other methods to ensure the robustness of our results. Using \texttt{pyaneti}, we ran similar fits to the one described above, but with: a) both planets on circular orbits; b) the inner planet on a circular orbit and the outer planet on an eccentric orbit; c) a fit with \texttt{pyaneti} but not multi-dimensional (i.e., fitting on only the RV data points); d) a joint model including both transit and RV data. Our joint fit of transit and RV data provides consistent results with the planet mass within 1~$\sigma$ of our other models (see Table~\ref{tab:resultscomp}).

We also make use of the radial velocity fitting toolkit \texttt{RadVel} \citep{fulton_radvel_2018}. Using \texttt{RadVel} we ran fits for: a) a two planet system with no GP to account for stellar activity; b) a 1 planet system (only the transiting inner planet) including a GP using the Celerite quasi-periodic kernel \citep{foreman-mackey_fast_2017}; c) a 2 planet system using the Celerite quasi-periodic kernel where both planets are on circular orbits; d) the same kernel again but where both planets are on eccentric orbits; e) a 2 planet system where we use the square-exponential GP kernel (described in \cite{fulton_radvel_2018}). As well as this, we checked for possible additional signals by fitting for a 3-planet system. We found that the third signal is unconstrained and the BIC and AIC increase slightly over the 2-planet case -- therefore we do not believe there are signs of additional planets in the system. We also changed the choice of priors in our models, in particular for the period of planet\,c to ensure we are not biasing our results. For example, the \texttt{RadVel} fit for a 2-planet system with the square exponential kernel has a uniform prior on $P_\mathrm{c}$ of 0 to 100 days, and the \texttt{RadVel} fit for a 2-planet system with the Celerite quasi-periodic kernel has a Gaussian prior of 50.6\,$\pm$\,1.0~days. In all cases the results are consistent. The resulting K semi-amplitude of TOI-544\,b found from all models are shown in Table~\ref{tab:resultscomp}.

\texttt{RadVel} also allows for model comparison. We found that in models which include more than one planet, a single-planet model is ruled out in every case - providing greater assurance of the existence of the second planet. Specifically, for a 1-planet system, the BIC is 1993.90 and AIC is 1965.46, and for 2-planet system 801.24 and 765.90 respectively. We also ran additional models using only the HARPS data points to ensure the HARPS-N points (which have a slightly higher level of scatter) are not influencing our fit, and with RVs extracted with {\tt TERRA} rather than {\tt SERVAL}, in all cases the results are consistent within 1-$\sigma$. As well as this, we ran \texttt{pyaneti} fits with different activity indicators:  FWHM, and contrast, and combinations of indicators: S-index and FWHM, S-index and contrast, and FWHM and contrast. In all cases the derived planet parameters are consistent with our other models within 1-$\sigma$, however the GP hyperparameters are not all so well constrained so we still use the model with the S-index as our adopted results.

\begin{table}
    \caption{Results of the transit and RV fit for TOI-544, showing the median value and 68 $\%$ credible interval for each parameter.}
    \label{tab:allresults}
    \begin{tabular}{ll}
    \toprule
    Fitted Parameter & Median value\\ \midrule
    Planet Transit Parameters\\ \midrule
    Stellar Mass $M_\star$ [$M_\odot$] & $0.631 \pm 0.018$ \\
    Stellar Radius $R_\star$ [$R_\odot$] & $0.623 \pm 0.012$ \\
    \noalign{\smallskip}
    Time of Mid-Transit $T_0$ [BJD$_\mathrm{TDB}-2450000$] & $9199.031363^{+0.000614}_{-0.000703}$ \\
    \noalign{\smallskip}
    Orbital Period $P_\mathrm{b}$ [days] & $1.548352 \pm 0.000002$ \\
    $R_\mathrm{b}/R_\star$ & $0.0297 \pm 0.0007$ \\
    Impact Parameter $b$ & $0.66^{+0.03}_{-0.04}$ \\
    \noalign{\smallskip}
    $\sigma_\mathrm{GP}$ [ppm] & $707^{+113}_{-86}$ \\
    \noalign{\smallskip}
    $\rho_\mathrm{GP}$ [hours] & $22.0^{+3.5}_{-2.8}$ \\
    $u_1$  & $0.43 \pm 0.16$ \\
    $u_2$  & $0.24 \pm 0.15$ \\
    $\log\sigma_\mathrm{S6}$ [ppt] & $-0.159094^{+0.005981}_{-0.005915}$  \\
    \noalign{\smallskip}
    $\log\sigma_\mathrm{S32}$ [ppt] & $-0.157482^{+0.005387}_{-0.005275}$ \\
    \noalign{\smallskip}
    Planet Radius $R_\mathrm{b}$ [$R_\oplus$] & \brad \\
    Semi-major Axis $a_\mathrm{b}$ [AU] & $0.0225 \pm 0.0002$ \\
    Equilibrium Temperature $T_\mathrm{eq}$ [K] & $999 \pm 14$ \\
    Transit Duration $T_{14}$ [hours] & $1.21 \pm 0.03$ \\
    \hline
    Planet RV Parameters \\\midrule
    Orbital Period,  $P_\mathrm{b}$ [days] &  $1.548352 \pm 0.000002$  \\
    Time of Inf. Conjunction, $T_\mathrm{conj, b}$ [BJD$_\mathrm{TDB}-2450000$] & 9199.0314 $\pm$ 0.0007  \\
    Time of Periastron, $T_\mathrm{peri, b}$  [BJD$_\mathrm{TDB}-2450000$] & 9198.6443 $\pm$ 0.0007\\
    Eccentricity,  $e_\mathrm{b}$ & $\equiv 0$  \\
    Argument of Periastron, $\omega_\mathrm{b}$ [$\degree$] & $\equiv 0$    \\
    $ew1_\mathrm{b}$, $\sqrt{e_\mathrm{b}}\sin\omega_\mathrm{b}$ & $\equiv 0$\\
    $ew2_\mathrm{b}$, $\sqrt{e_\mathrm{b}}\cos\omega_\mathrm{b}$ & $\equiv 0$\\
    RV Semi-Amplitude,  $K_\mathrm{b}$ [m s$^{-1}$] & $2.17\pm 0.36$    \\
    Planet Mass,  $M_\mathrm{b}$ [M$_{\oplus}$] & \bmass   \\
    \noalign{\smallskip}
    \noalign{\smallskip}
    \noalign{\smallskip}
    Orbital Period,  $P_\mathrm{c}$ [days] &  $50.089 \pm 0.24$  \\
    Time of Inf. Conjunction,  $T_\mathrm{conj, c}$ [BJD$_\mathrm{TDB}-2450000$] &  $9212.0^{+1.8}_{-1.9}$  \\
    \noalign{\smallskip}
    Time of Periastron $T_\mathrm{peri, c}$ [BJD$_\mathrm{TDB}-2450000$] & $9205.4^{+2.2}_{-2.8}$  \\
    \noalign{\smallskip}
    Eccentricity,  $e_\mathrm{c}$ &  $0.32^{+0.08}_{-0.09}$  \\
    \noalign{\smallskip}
    Argument of Periastron,  $\omega_\mathrm{c}$ [$\degree$] & $11^{+ 19}_{-17}$   \\
    $ew1_\mathrm{c}$, $\sqrt{e_\mathrm{c}}\sin\omega_\mathrm{c}$ & $0.11 \pm 0.17$ \\
    $ew2_\mathrm{c}$, $\sqrt{e_\mathrm{c}}\cos\omega_\mathrm{c}$ & $0.52^{+0.08}_{-0.11}$ \\
    RV Amplitude, $K_\mathrm{c}$ [m s$^{-1}$ ]  & $5.36\pm0.56$   \\
    Planet Minimum Mass,  $M_\mathrm{c} \sin i_\mathrm{c}$ [M$_{\oplus}$] & \cmass \\
    \midrule
    Other Parameters \\ 
    \midrule
    Offset RV$_\mathrm{HARPS-N}$ [km\,s$^{-1}$]  & $-0.016 \pm 0.002$   \\
    Offset RV$_\mathrm{HARPS}$ [km\,s$^{-1}$] & $0.006 \pm 0.001$  \\
    Offset S-index$_\mathrm{HARPS-N}$  &  $1.13 \pm 0.03$ \\
    Offset S-index$_\mathrm{HARPS}$  &  $1.25^{+0.03 }_{-0.04}$   \\
    \noalign{\smallskip}
    Jitter Term RV$_\mathrm{HARPS-N}$,  $\sigma_{\rm HARPS-N}$ [m\,s$^{-1}$] & $2.75^{+1.44}_{-1.13}$  \\
    \noalign{\smallskip}
    Jitter Term RV$_\mathrm{HARPS}$,  $\sigma_{\rm HARPS}$ [m\,s$^{-1}$] & $1.85^{+ 0.36}_{-0.33}$ \\
    \noalign{\smallskip}
    Jitter Term S-index$_\mathrm{HARPS-N}$  & $45.5^{+17.7}_{-12.8}$  \\
    \noalign{\smallskip}
    Jitter Term S-index$_\mathrm{HARPS}$  & $42.6^{+3.5}_{-3.2}$ \\
    \midrule
    GP Hyperparameters \\ 
    \midrule
    $A_0$ &  $0.002^{+0.001}_{-0.001}$   \\
    \noalign{\smallskip}
    $A_1$ &  $0.025^{+0.008}_{-0.005}$  \\
    \noalign{\smallskip}
    $A_2$ &  $0.076^{+0.025}_{-0.016}$  \\
    $A_3$ & $\equiv 0$  \\
    $\lambda_\mathrm{e}$ [days] &   $112^{+28}_{-29}$  \\
    \noalign{\smallskip}
    $\lambda_\mathrm{p}$  &   $0.519^{+0.091}_{- 0.074 }$  \\
    \noalign{\smallskip}
    Rotation Period,  P$_\mathrm{GP}$ [days] &  $19.343^{+0.073}_{-0.076}$  \\ \bottomrule
    \end{tabular}
\end{table}

\section{Discussion}
\subsection{Composition of TOI-544\texorpdfstring{\,b} {}}
\label{composition}

\begin{figure*}
	\includegraphics[width=1.9\columnwidth]{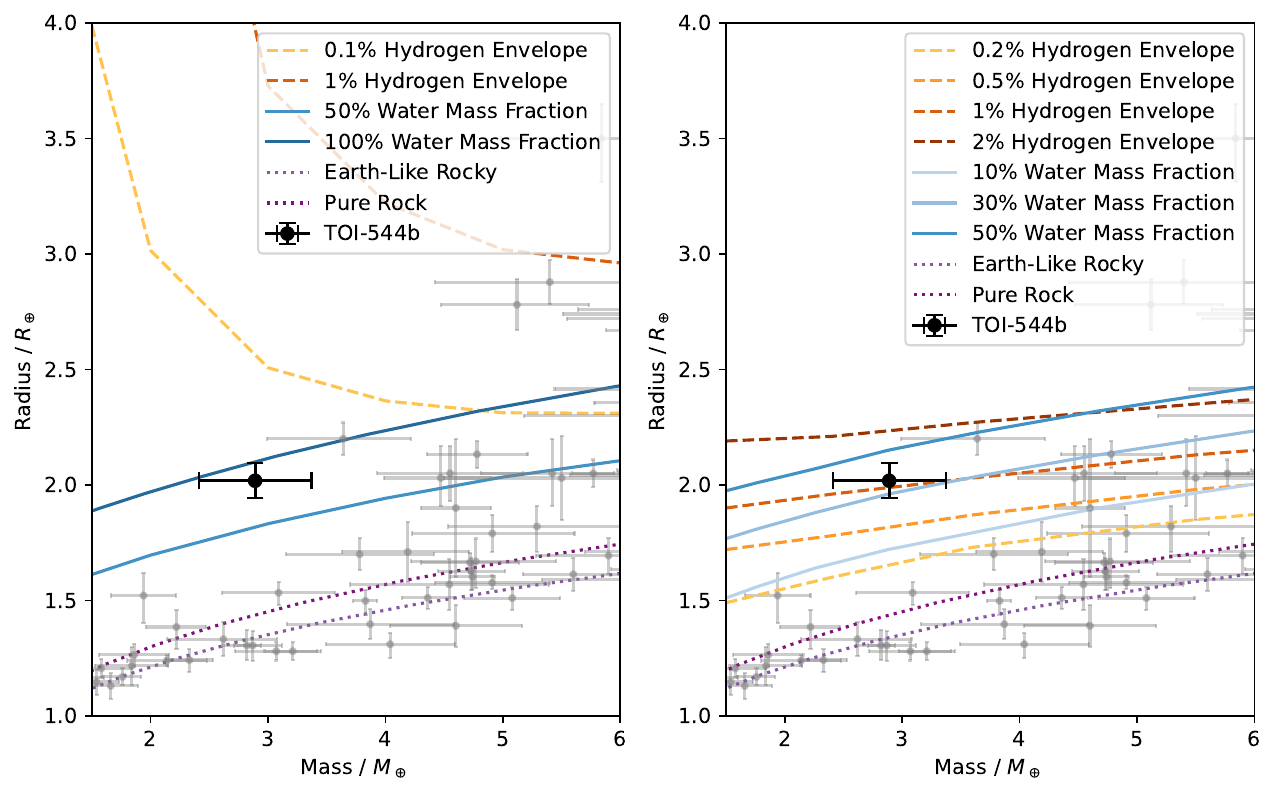}
    \caption{Mass-radius diagram showing confirmed planets with mass uncertainties < 20$\%$ in grey circles, with TOI-544\,b shown as the black circle. Left: the coloured tracks show different potential planet compositions from \protect\cite{Zeng2019}, assuming a planetary equilibrium temperature of 1000 K. The dashed lines show compositions of an Earth-like rocky core surrounded by a layer of H-He in varying percentages by mass, the solid lines show water world compositions with varying water mass fractions, the dotted lines show rocky and Earth-like rocky compositions. TOI-544\,b is closest to the tracks with 100$\%$ H$_2$O and 50$\%$ H$_2$O with 50$\%$ rock. Right: the same as the left panel but the the dashed lines show compositions of an Earth-like rocky core surrounded by a layer of H-He in varying percentages by mass from \protect\cite{LopFor14} using the models for a 10Gyr planet, with solar metalicity and flux of 10F$_{\oplus}$, the solid lines show compositions from \protect\cite{Aguichine2021} of irradiated ocean worlds with varying water mass fractions, and the dotted lines show the same rocky and Earth-like rocky compositions from \protect\cite{Zeng2019}. TOI-544\,b is closest to models with either 30$\%$ water mass fraction, or a rocky core with a Hydrogen envelope of between 0.5 - 1$\%$. The data is downloaded from the NASA Exoplanet Archive.}
    \label{fig:toi544_mr}
\end{figure*}

The calculated density of TOI-544\,b is $ \rho_\mathrm{b}$\,=\,\bdensity~g~cm$^{-3}$, which is not dense enough to be composed of entirely rock and iron, where densities are typically between 3 g~cm$^{-3}$ and 10 g~cm$^{-3}$ \citep{Zeng2019}. This implies that a composition of a bare rocky-iron core can likely be excluded.  TOI-544\,b must have some additional component to its composition, but whether it is in the form of water/ice layers or a H-He atmosphere depends on which particular composition models are used. Figure \ref{fig:toi544_mr}, left panel, shows the mass radius diagram for small planets (< 4 \rearth, with mass uncertainties < 20 $\%$) with TOI-544\,b highlighted and composition tracks from \cite{Zeng2019} shown in different colours. As noted in \cite{rogers2023}, these models are often misinterpreted when used in mass-radius diagrams. Specifically, there are a range of models available depending on the chosen temperature, this temperature is often assumed to be the equilibrium temperature of the planets but is actually the temperature at a pressure of 100 bars. In this case we have used the {\textit{incorrect}} temperature (equal to the planet equilibrium temperature of $\approx$ 1000 K) in order to show a direct comparison to many other mass-radius diagrams available in the literature. Using these models alone, it appears that TOI-544\,b does not fit the super-Earth (rocky/iron core) scenario, and that it also does not fit the typical sub-Neptune composition of a rocky core surrounded by atmospheric H-He. From these models it seems likely that TOI-544\,b is a water-world planet with a sizeable fraction of H$_2$O present. From the \cite{Zeng2019} models it seems that a water fraction by mass of over 50$\%$ is possible for this planet. There are few other planets detected in a similar area of parameter space, however we note that many of the planets in this regime have multiple mass values listed on the NASA Exoplanet Archive so their exact location on the mass-radius diagram depends on the specific choice of literature parameters, and many have longer orbital periods, meaning it is difficult to do a real comparison.

To further investigate the compositional nature of the planet, we use the dimensionless parameter $\zeta$ from \cite{zeng2021}
\begin{equation}
\zeta \equiv \frac{(R_c/R_{\oplus})}{(M_c/M_{\oplus})^{0.25}},
\end{equation}
where $R_C$ and $M_C$ are the core radius and mass respectively, similar to the approach taken in \cite{nava_exoplanet_2019}. This parameter can be used to distinguish between the 3 possible compositions of small exoplanets -- either rocky and earth-like ($\zeta = 1$), having significant amounts of ices ($\zeta = 1.4$), or icey cores with hydrogen/helium envelopes ($\zeta = 2.2$). For small planets the core radius and mass can be approximated the planet radius and mass (in units of Earth radii and mass). The value of $\zeta$ for TOI-544\,b is 1.5, which suggests an ice dominated composition.

Another approach is noting that TOI-544 b is a highly irradiated planet due to its short semi-major axis. We make use of the \texttt{Structure Model INterpolator (SMINT)}\footnote{\url{https://github.com/cpiaulet/smint}} which obtains posterior distributions of H$_2$O mass fraction based on interpolation onto the \cite{Aguichine2021} model grids. Using the stellar and planetary parameters listed previously, with conservative uncertainties, we obtain an H$_2$O fraction of 0.25 $\pm$ 0.12. 
This is lower than the expected H$_2$O fraction seen in the mass-radius diagram. We note, however, that the models from \cite{Zeng2019} do not account for the high level of radiation such a close-in planet would receive. 

However, whilst \cite{Zeng2019} models are commonly used in mass-radius diagrams for small planets, \cite{rogers2023} recommend using the mass-radius relations for sub-Neptune compositions given in \cite{LopFor14}, which assume a constant planet age, rather than the \cite{Zeng2019} models which assume a constant specific entropy. In the right panel of Figure \ref{fig:toi544_mr}, we plot a mass-radius diagram with composition tracks from \cite{LopFor14} and \cite{Aguichine2021}. The dashed lines show compositions of an Earth-like rocky core surrounded by a layer of H-He in varying percentages by mass from \cite{LopFor14}. The solid lines show compositions from \cite{Aguichine2021} of irradiated ocean worlds with varying water mass fractions. The dotted lines show the same rocky and Earth-like rocky compositions from \cite{Zeng2019} which are shown in the left panel. Setting TOI-544\,b on this graph, we see that according to the \cite{LopFor14} models, a composition of an Earth-like core surrounded by a layer of H-He of between 0.5 and 1$\%$ by mass can also explain the observed mass and radius. The composition tracks from \cite{Aguichine2021} for irradiated water-worlds suggest an alternative composition of TOI-544\,b of a rocky core with a layer of water/ice of around 30$\%$ by mass. 

As discussed in \cite{rogers2023}, for individual planets such as TOI-544\,b, mass and radius alone are insufficient to uniquely constrain planet composition. In order to break the degeneracy between the water-worlds and sub-Neptune models, atmospheric observations of the planet are needed to rule out (or in) the potential for H-He or H$_2$O atmospheres. Fortunately, as described in Section \ref{atmospherics}, TOI-544 b is an ideal candidate for atmospheric studies. Future observations, with for instance JWST, should be able to help determine more definitively whether this is a water-world or not. 

From Figure \ref{fig:toi544_mr} it can also be seen that TOI-544\,b sits within an area of the mass-radius diagram where few planets have been observed. Of the more than 5000 exoplanets confirmed to date, there are less than 200 small (< 4 R$_{\oplus}$) planets which have precisely characterised masses (uncertainties < 20$\%$), and of those, less than 30 are low mass (< 4M$_{\oplus}$). The majority of these small, low mass planets with precise characterisation cluster around the composition tracks for rocky silicates or iron composition -- similar to an Earth-like composition. The most similar planet is Kepler-307c, which is one of the only low mass planets that has a similarly low density (seen to the top right of TOI-544\,b in Figure \ref{fig:toi544_mr}). The mass of Kepler-307c was determined through transit timing variations rather than through RV observations. As there have been suggestions of a potential offset between the two methods \citep{steffen_sensitivity_2016, millsmazeh}, it is possible that these two planets are not fully comparable. In general, there are few precisely-characterised masses for small planets, and even fewer for potential water-worlds, meaning TOI-544\,b is an important addition to this region of parameter space.

\subsection{Location in relation to the radius valley}
\label{radius valley}

\begin{figure}
	\includegraphics[width=\columnwidth]{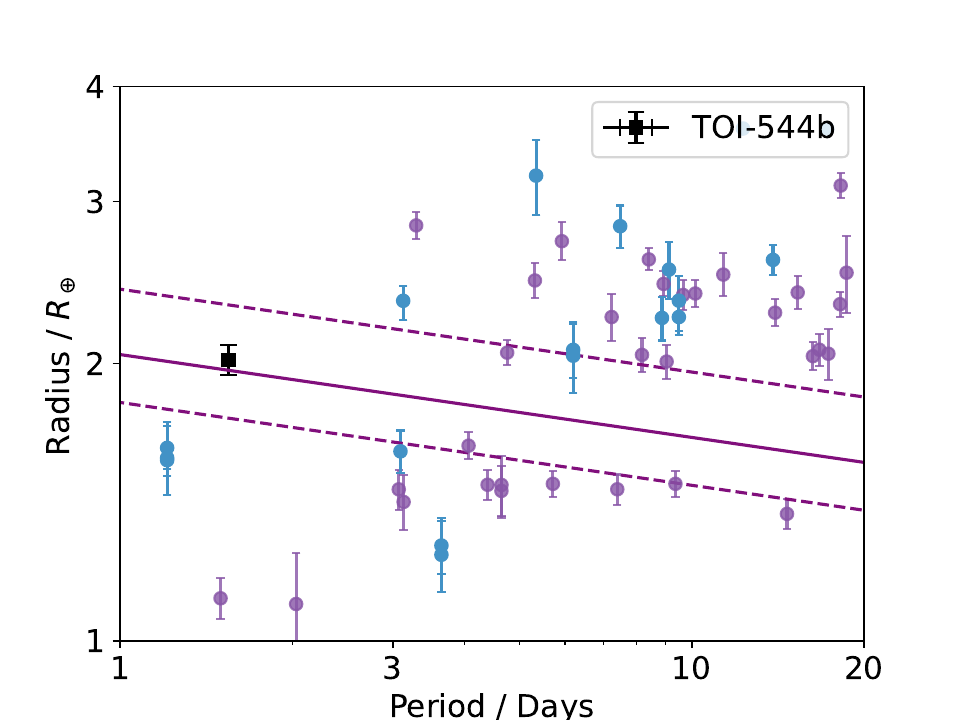}
    \caption{Period-radius diagram for a subset of confirmed planets orbiting stars with masses < 0.8 M$_{\odot}$. Purple circles are confirmed Kepler planets fitted in a homogeneous way in \protect\cite{ho_deep_2023}. Blue circles show planets with precise mass measurements from the NASA Exoplanet Archive. TOI-544\,b shown by the black square. TOI-544\,b sits within the expected location of the radius valley (between the dashed lines) for this specific stellar mass - calculated from Equation 11 in \protect\cite{ho_deep_2023}.}
    \label{fig:radiusvalley}
\end{figure}

In the most recent comprehensive study of the radius valley, \cite{ho_deep_2023} refitted Kepler data to find an empirical radius valley location, as a function of various other parameters. In particular, they find a dependence on the location of the valley as a function of stellar mass. Figure \ref{fig:radiusvalley} shows the period-radius diagram for confirmed Kepler planets orbiting stars with stellar mass $<$ 0.8 M$_{\odot}$ fitted in a homogeneous way in \cite{ho_deep_2023}. 
Additionally, planets with precise mass measurements from the NASA Exoplanet Archive which orbit stars with masses $<$ 0.8 M$_{\odot}$ are highlighted, and TOI-544\,b shown. 
TOI-544\,b sits within the radius valley region calculated for this specific stellar mass - using Equation 11 from \citep{ho_deep_2023}:
\begin{equation}
\label{equationvalley}
    \log_{10}(R_p/R_{\oplus}) = A\,\log_{10}(P/days) + B\,\log_{10}(M_*/M_{\odot}) + C
\end{equation}
with A = -0.09$^{+0.02}_{-0.03}$, B = 0.21$^{+0.06}_{-0.07}$, C = 0.35$^{+0.02}_{-0.02}$ and using TOI-544's stellar mass (see Table~\ref{Table:stellarparameters}).
TOI-544\,b is more than $3\sigma$ away from the upper and lower bounds of the radius valley (shown by the dashed lines). We also ran a number of fits for stellar radius; we use the calculated stellar radii from \ref{Table:stellarparameters}, taking the most extreme cases of the BASTA fit -1$\sigma$, and the astroARIADNE fit +1$\sigma$ (i.e.the smallest and largest possible from our results), we find planet radii of 2.16 and 1.98 \rearth respectively. These values still put TOI-544 within the limits of the valley given in \cite{ho_deep_2023}.

We also compare the location of the radius valley presented in \cite{ho_deep_2023} with other works. In particular, \cite{petigura_california-kepler_2022} also find that the location of the valley varies based on stellar mass, finding a similar relation to Equation \ref{equationvalley}. For a star of mass 0.5 - 0.7 \Msun, Figure 8 in \cite{petigura_california-kepler_2022} shows that TOI-544\,b would be inside the valley. \cite{Cloutier2020} also investigated the location of the radius valley for low mass (mid-K to mid-M dwarf) stars and similarly find that a planet such as TOI-544\,b, with radius of \brad \rearth\ and orbital period of 1.55 days would be located inside the valley in the region they dub ``keystone planets'', see Figure 15 of \cite{Cloutier2020}.

The observational results for the location of the radius valley can be compared with theoretical models such as in \cite{owen_evaporation_2017}, which predicts that the location of the valley also depends on the planetary core composition. If we assume that planets form uniformly with an icey core -- rather than a rocky one -- then the theoretical models predict that the radius valley will be shifted to a higher radius for a given orbital period. This means that, for planets with icey cores - sometimes referred to as water world planets - the location of TOI-544\,b in the radius-period space would not put it inside the radius valley. Instead it would sit below the valley, in a region which predicts planets with a sizeable water/ice faction but without significant hydrogen/helium atmospheres, potentially having undergone atmospheric loss. 

\cite{luque_density_2022} argue that the radius valley distribution may in fact be the result of two different core compositions of planets rather than purely from atmospheric loss mechanisms, in the case of planets orbiting M dwarfs. They state that small planets come in two distinct types: super-Earths with rocky/iron composition, and water worlds with a combination of both rock and water/ices. This interpretation is disputed by \cite{rogers2023}, who argue that the properties of the sample of planets around M dwarfs studied by \cite{luque_density_2022} can also be explained by the more traditional super-Earth sub-Neptune classifications which arise from atmospheric loss models.

We note that at this stage there is no conclusive evidence either way to support the water worlds versus atmospheric loss explanation of the radius valley, and further investigation is needed to fully distinguish between the two theories. In particular, confirmation of TOI-544\,b as a water world or not (as well as other small planets) would help to provide evidence for the formation mechanisms which carve the radius valley.

\subsection{Planet c}
We searched the TESS lightcurves for signs of a transit of TOI-544\,c but none were found. If we assume that both planets are at the same orbital inclination (which may not be the case) then planet\,c would not be expected to transit given its impact parameter of ~5. As a result, we can only constrain a minimum mass for TOI-544\,c of \cmass  $M_{\oplus}$, for reference this is slightly higher than the mass of Neptune. Planet c is found to have a non-zero eccentricity, and, if it is confirmed that TOI-544\,b has a large fraction of water within its composition, then it is likely that it must have formed exterior to the snow-line, and then migrated inwards - this migration could have been facilitated by TOI-544\,c. A full dynamical investigation of the system architecture is beyond the scope of this paper, but could be interesting in the future, particularly if the composition of the inner planet is better constrained, and it is possible that if the planets do not orbit in the same plane then plant\,c could in fact transit.

\subsection{Potential atmospheric studies of TOI-544\texorpdfstring{\,b} {}}
\label{atmospherics}

\begin{figure}
	\includegraphics[width=\columnwidth]{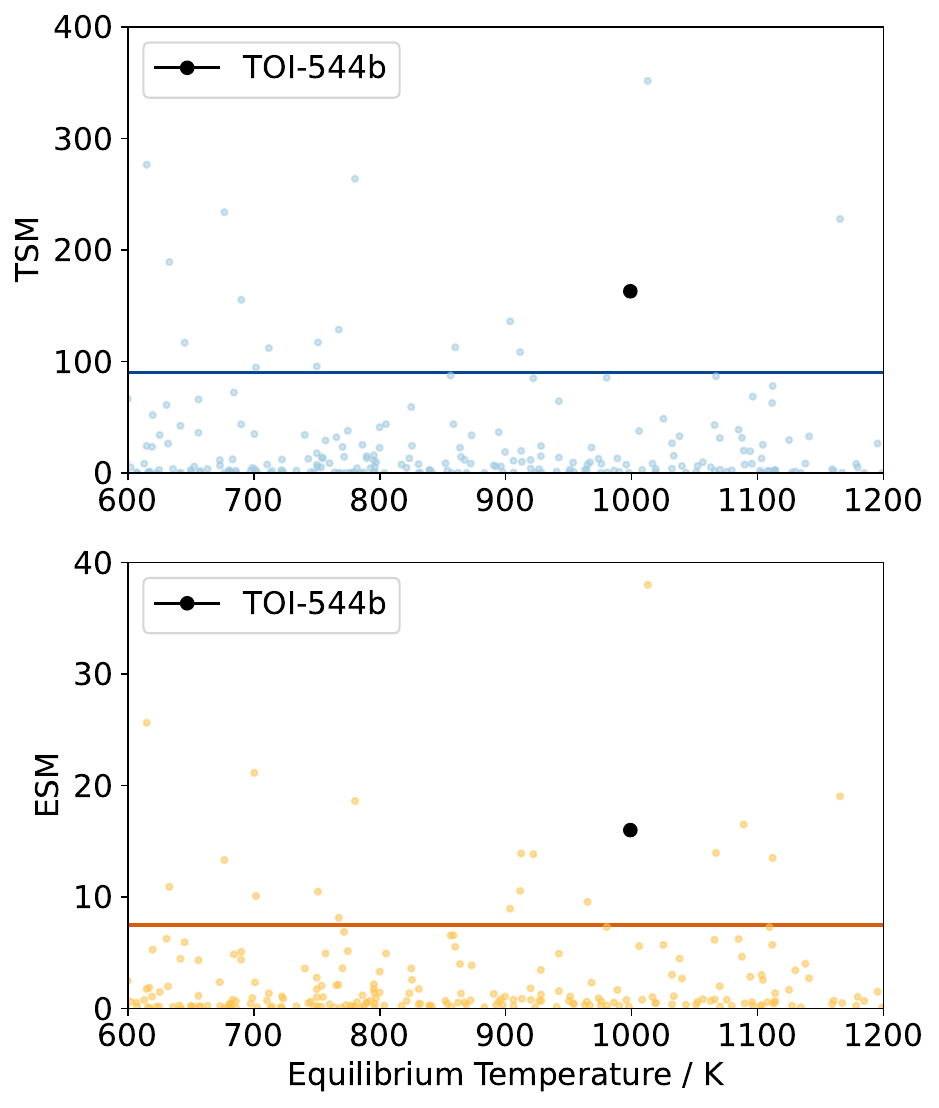}
    \caption{Top panel: The transmission spectroscopy metric (TSM) against equilibrium temperature of all small (R $< 4 R_{\oplus}$) confirmed planets with a mass measurement, blue dots. The threshold given in \protect\cite{Kempton18} is shown by the solid line. TOI-544\,b is shown in the black circle, it is within the top 15 planets for TSM value. Bottom Panel: Same as above but for the emission spectroscopy metric (ESM). TOI-544\,b is within the top 10 planets for ESM value. The data is downloaded from the NASA Exoplanet Archive.}
    \label{fig:tsmesm}
\end{figure}

We calculated the transmission spectroscopy metric \citep[TSM,][]{Kempton18} for TOI-544\,b, finding a value of 163, much higher than the recommended threshold value (90) from \citep{Kempton18} for planets with radii above 1.5 R$_{\oplus}$, placing TOI-544\,b among the most appealing targets for transmission spectroscopy characterisation with JWST (within the top 15 targets for similar planets, see Figure \ref{fig:tsmesm}). Moreover, the emission spectroscopic metric \citep[ESM,][]{Kempton18} is found to be 16. Considering the \cite{Kempton18} cutoff of 7.5, TOI-544\,b lies among the top 10 most favourable targets (see Figure \ref{fig:tsmesm}). TOI-544\,b is within the top 5 targets for both TSM and ESM for planets with radii between 1.5 and 2.75 R$_{\oplus}$ and temperature between 800 and 1250K, as identified by the TESS follow-up atmospheric working group.  We also calculate the predicted S/N ratio for atmospheric observations of TOI-544\,b with JWST (using the method in \cite{niraula2017}, finding a value of 1.000. This puts TOI-544\,b within the top dozen small planets with temperature less than 1250K in terms of potential JWST observations. Atmospheric observations should help to reduce the degeneracy between composition models for this planet and determine whether a water world or a rocky and hydrogen composition is more likely, similar to a recent study of TOI-270\,d \citep{VanEylen21} for which transmission spectroscopy revealed a possible hydrogen-rich atmosphere \citep{mikalevans2022}.

\section{Conclusions}
\label{conclusions}

We present the results of an extensive high-precision RV campaign of TOI-544. We confirm the planet TOI-544\,b and derive a mass of $M_\mathrm{b}$\,=\,\bmass~M$_{\oplus}$ which, combined with the planetary radius of $R_\mathrm{b}$\,=\,\brad R$_{\oplus}$ gives a bulk density of $\rho_\mathrm{b}$\,=\,\bdensity~g~cm$^{-3}$. The density of the planet means it most likely has either a significant fraction of ice within its composition (around 30$\%$ by mass) or is composed of an Earth-like rocky core surrounded by a layer of atmospheric H-He (around 0.5 - 1 $\%$ by mass). TOI-544\,b also sits within the expected location of the small planet radius valley for FGK stars, although improvements in the radius measurement with additional transit observations would help confirm this further. The calculated TSM and ESM of TOI-544\,b put it within the top few planets for atmospheric observations with similar size and temperature, meaning it is an excellent candidate for future observations with JWST. We additionally confirm the existence of a second, non-transiting planet within the system, TOI-544\,c, with a minimum mass of $M_\mathrm{c} \sin i_\mathrm{c}$\,=\,\cmass M$_{\oplus}$. Both planets have well-characterised masses (uncertainties of < 20$\%$) and contribute to the small but growing number of small planets with precisely characterised masses.

\section*{Acknowledgements}

We thank the anonymous reviewer for reviewing the paper, and for their valuable suggestions which have improved the manuscript.
This work is done under the framework of the KESPRINT collaboration (http://kesprint.science). KESPRINT is an international consortium devoted to the characterization and research of exoplanets discovered with space-based missions. Based on observations made (a) with ESO-3.6 m telescope at La Silla Observatory (Chile) under programme IDs 106.21TJ.001, 0103.C-0442, and 60.A-9709; (b) with the Italian Telescopio Nazionale Galileo (TNG) operated on the island of La Palma in the Spanish Observatorio del Roque de los Muchachos (ORM) by the INAF-Fundaci\'on Galileo Galilei, under programme IDs CAT19A\_162 and CAT19A\_97. We are extremely grateful to the ESO and TNG staff members for their unique and superb support during the observations. We are grateful to Cynthia Ho for discussions on the small planet radius valley and James Owen for discussion on use of planet composition tracks. HLMO would also like to thank the Science and Technology Facilities Council (STFC) for funding support through a PhD studentship. CMP gratefully acknowledge the support of the  Swedish National Space Agency (DNR 65/19). GN thanks for the research funding from the Ministry of Education and Science programme the "Excellence Initiative - Research University" conducted at the Centre of Excellence in Astrophysics and Astrochemistry of the Nicolaus Copernicus University in Toru\'n, Poland. RL acknowledges funding from University of La Laguna through the Margarita Salas Fellowship from the Spanish Ministry of Universities ref. UNI/551/2021-May 26, and under the EU Next Generation funds. APH and ME acknowledge the support of DFG Grant HA 3279/12-1 within the DFG Schwerpunkt SPP 1992, Exploring the Diversity of Extrasolar Planets. HJD acknowledges support from the Spanish Research Agency of the Ministry of Science and Innovation (AEI-MICINN) under grant PID2019-107061GB-C66, DOI: 10.13039/501100011033. KWFL was supported by Deutsche Forschungsgemeinschaft grants RA714/14-1 within the DFG Schwerpunkt SPP 1992, Exploring the Diversity of Extrasolar Planets. JSJ greatfully acknowledges support by FONDECYT grant 1201371 and from the ANID BASAL project FB210003.JK gratefully acknowledges the support of the Swedish National Space Agency (SNSA; DNR 2020-00104) and of the Swedish Research Council  (VR: Etableringsbidrag 2017-04945).

\section*{Data Availability}

This paper includes raw data collected by the TESS mission, which are publicly available from the Mikulski Archive for Space Telescopes (MAST, https://archive.stsci.edu/tess). Some of the observations made with HARPS at the ESO 3.6m telescope (programme 0103.C-0442(A)) are publicly
available at the ESO archive (http://archive.eso.org/), observations which are not currently available will become open access within the next year. All processed data underlying this article (including observations from HARPS-N) are available in the article and in its online supplementary material.


\bibliographystyle{mnras}
\bibliography{MyLibrary}



\appendix
\section{Appendix}
\label{Appendix}
\onecolumn

\begin{small}
\centering
\tablefirsthead{
\hline
\hline
\multicolumn{1}{c}{BJD$_\mathrm{TBD}$} &
\multicolumn{1}{c}{RV} &
\multicolumn{1}{c}{$\sigma_\mathrm{RV}$} &
\multicolumn{1}{c}{FWHM} &
\multicolumn{1}{c}{BIS} &
\multicolumn{1}{c}{Contrast} &
\multicolumn{1}{c}{$\mathrm{T_{exp}}$} & 
\multicolumn{1}{c}{SNR} \\
\multicolumn{1}{c}{$-$2450000} &
\multicolumn{1}{c}{($\mathrm{km\,s^{-1}}$)} &
\multicolumn{1}{c}{($\mathrm{km\,s^{-1}}$)} &
\multicolumn{1}{c}{($\mathrm{km\,s^{-1}}$)} &
\multicolumn{1}{c}{($\mathrm{km\,s^{-1}}$)} &
\multicolumn{1}{c}{(\%)} &
\multicolumn{1}{c}{(s)} &
\multicolumn{1}{c}{@550nm} \\
\hline
}
\tablehead{
\multicolumn{8}{c}
{{\bfseries \tablename\ \thetable{}} -- continued from previous page.}\\
\hline
\hline
\multicolumn{1}{c}{BJD$_\mathrm{TBD}$} &
\multicolumn{1}{c}{RV} &
\multicolumn{1}{c}{$\sigma_\mathrm{RV}$} &
\multicolumn{1}{c}{FWHM} &
\multicolumn{1}{c}{BIS} &
\multicolumn{1}{c}{Contrast} &
\multicolumn{1}{c}{$\mathrm{T_{exp}}$} & 
\multicolumn{1}{c}{SNR} \\
\multicolumn{1}{c}{$-$2450000} &
\multicolumn{1}{c}{($\mathrm{km\,s^{-1}}$)} &
\multicolumn{1}{c}{($\mathrm{km\,s^{-1}}$)} &
\multicolumn{1}{c}{($\mathrm{km\,s^{-1}}$)} &
\multicolumn{1}{c}{($\mathrm{km\,s^{-1}}$)} &
\multicolumn{1}{c}{(\%)} &
\multicolumn{1}{c}{(s)} &
\multicolumn{1}{c}{@550nm} \\
\hline
}
\tabletail{
\hline
\multicolumn{8}{c}{{Continued on next page}}\\
\hline
}
\tablelasttail{
\hline
}
\tablecaption{Absolute radial velocities and spectral activity indicators measured from the HARPS spectra with the {\tt DRS}.
\label{table-TOI-0544-3p6_harps-0108-drs-complete_output}}
\begin{supertabular}{cccccccc}
   8578.51514 &   8.3561 &   0.0035 &   6.5993 &   0.0833 &   38.785 & 1800 &  35.5 \\
   8579.53150 &   8.3449 &   0.0030 &   6.6112 &   0.0635 &   38.743 & 1500 &  40.8 \\
   9205.63053 &   8.3640 &   0.0026 &   6.4372 &   0.0499 &   39.787 & 1800 &  43.0 \\
   9206.63247 &   8.3729 &   0.0018 &   6.4598 &   0.0446 &   39.642 & 1800 &  58.6 \\
   9207.57568 &   8.3665 &   0.0022 &   6.4685 &   0.0398 &   39.526 & 1800 &  48.5 \\
   9207.66046 &   8.3655 &   0.0018 &   6.4766 &   0.0470 &   39.529 & 1800 &  58.1 \\
   9214.71815 &   8.3418 &   0.0019 &   6.4361 &   0.0721 &   39.454 & 2400 &  57.0 \\
   9215.64475 &   8.3462 &   0.0017 &   6.4391 &   0.0708 &   39.640 & 2400 &  62.0 \\
   9217.68640 &   8.3486 &   0.0019 &   6.4223 &   0.0515 &   39.845 & 1800 &  55.1 \\
   9218.68225 &   8.3483 &   0.0017 &   6.4289 &   0.0527 &   39.817 & 1800 &  61.0 \\
   9219.65704 &   8.3470 &   0.0016 &   6.4178 &   0.0542 &   39.848 & 1800 &  64.6 \\
   9221.60896 &   8.3482 &   0.0017 &   6.4250 &   0.0660 &   39.860 & 1800 &  60.8 \\
   9222.61015 &   8.3382 &   0.0016 &   6.4123 &   0.0583 &   39.892 & 1800 &  64.0 \\
   9223.56895 &   8.3505 &   0.0017 &   6.4189 &   0.0525 &   39.943 & 1800 &  60.7 \\
   9224.66129 &   8.3476 &   0.0018 &   6.4348 &   0.0403 &   39.820 & 1800 &  58.5 \\
   9226.65544 &   8.3572 &   0.0019 &   6.4612 &   0.0390 &   39.638 & 1800 &  56.1 \\
   9227.66789 &   8.3493 &   0.0023 &   6.4623 &   0.0571 &   39.592 & 1800 &  46.5 \\
   9228.64010 &   8.3423 &   0.0028 &   6.4661 &   0.0659 &   39.562 & 1800 &  40.5 \\
   9230.63045 &   8.3425 &   0.0017 &   6.4496 &   0.0663 &   39.652 & 1800 &  61.3 \\
   9231.63611 &   8.3439 &   0.0025 &   6.4623 &   0.0673 &   39.691 & 1800 &  44.2 \\
   9232.63467 &   8.3486 &   0.0016 &   6.4501 &   0.0619 &   39.705 & 1800 &  64.8 \\
   9233.67770 &   8.3443 &   0.0018 &   6.4614 &   0.0573 &   39.736 & 1800 &  59.0 \\
   9242.62501 &   8.3451 &   0.0016 &   6.4328 &   0.0650 &   39.495 & 2400 &  67.8 \\
   9243.56834 &   8.3522 &   0.0017 &   6.4395 &   0.0521 &   39.540 & 1800 &  62.0 \\
   9244.54494 &   8.3552 &   0.0019 &   6.4387 &   0.0528 &   39.538 & 1800 &  55.9 \\
   9244.62785 &   8.3553 &   0.0017 &   6.4513 &   0.0519 &   39.516 & 1800 &  63.9 \\
   9245.60955 &   8.3526 &   0.0027 &   6.4711 &   0.0454 &   39.245 & 1800 &  41.7 \\
   9246.63208 &   8.3594 &   0.0020 &   6.4711 &   0.0625 &   39.388 & 1800 &  53.8 \\
   9247.59527 &   8.3478 &   0.0021 &   6.4614 &   0.0628 &   39.417 & 1800 &  50.9 \\
   9248.61373 &   8.3503 &   0.0032 &   6.4636 &   0.0792 &   39.252 & 2700 &  36.4 \\
   9249.62657 &   8.3502 &   0.0022 &   6.4572 &   0.0698 &   39.648 & 2100 &  49.9 \\
   9250.57953 &   8.3420 &   0.0027 &   6.4415 &   0.0538 &   39.744 & 1860 &  42.2 \\
   9251.59236 &   8.3586 &   0.0026 &   6.4425 &   0.0492 &   39.694 & 1800 &  43.7 \\
   9256.62889 &   8.3508 &   0.0021 &   6.4400 &   0.0542 &   39.724 & 1800 &  53.8 \\
   9257.58116 &   8.3564 &   0.0017 &   6.4369 &   0.0595 &   39.724 & 1800 &  64.6 \\
   9261.63836 &   8.3525 &   0.0031 &   6.4272 &   0.0582 &   39.488 & 1800 &  38.4 \\
   9262.59334 &   8.3549 &   0.0028 &   6.4447 &   0.0652 &   39.466 & 1800 &  41.8 \\
   9264.60899 &   8.3479 &   0.0019 &   6.4604 &   0.0502 &   39.435 & 1800 &  59.2 \\
   9265.56638 &   8.3532 &   0.0019 &   6.4639 &   0.0633 &   39.396 & 1800 &  57.8 \\
   9266.55658 &   8.3535 &   0.0022 &   6.4603 &   0.0633 &   39.334 & 1800 &  50.4 \\
   9267.55362 &   8.3425 &   0.0020 &   6.4707 &   0.0640 &   39.352 & 1800 &  55.6 \\
   9269.57020 &   8.3487 &   0.0024 &   6.4821 &   0.0541 &   39.077 & 1800 &  47.9 \\
   9272.57672 &   8.3419 &   0.0021 &   6.4640 &   0.0555 &   39.208 & 1800 &  55.5 \\
   9273.59284 &   8.3397 &   0.0034 &   6.4738 &   0.0650 &   38.892 & 1800 &  37.2 \\
   9274.58962 &   8.3417 &   0.0022 &   6.5579 &   0.0610 &   38.735 & 1800 &  52.0 \\
   9275.56486 &   8.3413 &   0.0023 &   6.5514 &   0.0540 &   38.866 & 1800 &  49.6 \\
   9276.59246 &   8.3501 &   0.0026 &   6.5467 &   0.0656 &   38.799 & 1800 &  45.8 \\
   9277.57256 &   8.3518 &   0.0024 &   6.5421 &   0.0557 &   38.822 & 1800 &  48.9 \\
   9284.57306 &   8.3475 &   0.0024 &   6.5636 &   0.0653 &   38.916 & 1800 &  48.1 \\
   9287.53172 &   8.3509 &   0.0021 &   6.5785 &   0.0534 &   38.835 & 1800 &  54.6 \\
   9288.53857 &   8.3512 &   0.0022 &   6.5767 &   0.0612 &   38.844 & 1800 &  51.3 \\
   9290.53613 &   8.3509 &   0.0024 &   6.5842 &   0.0594 &   38.662 & 1800 &  48.7 \\
   9291.52201 &   8.3473 &   0.0019 &   6.6084 &   0.0724 &   38.602 & 1800 &  58.6 \\
   9294.53123 &   8.3411 &   0.0042 &   6.4445 &   0.0656 &   39.689 & 1800 &  31.0 \\
   9295.53344 &   8.3511 &   0.0021 &   6.5290 &   0.0583 &   39.035 & 2100 &  52.6 \\
   9296.51173 &   8.3565 &   0.0018 &   6.5513 &   0.0555 &   38.968 & 1800 &  61.4 \\
   9297.52848 &   8.3549 &   0.0019 &   6.5676 &   0.0610 &   38.844 & 1800 &  60.2 \\
   9490.84461 &   8.3381 &   0.0031 &   6.4877 &   0.0455 &   39.551 & 2100 &  35.6 \\
   9491.81716 &   8.3320 &   0.0034 &   6.4756 &   0.0611 &   39.570 & 2100 &  33.3 \\
   9497.84874 &   8.3623 &   0.0024 &   6.5573 &   0.0357 &   39.088 & 2100 &  45.0 \\
   9501.84420 &   8.3521 &   0.0028 &   6.6205 &   0.0841 &   38.682 & 2100 &  41.3 \\
   9502.76982 &   8.3440 &   0.0021 &   6.5560 &   0.0797 &   38.988 & 2100 &  50.3 \\
   9503.79777 &   8.3469 &   0.0033 &   6.5105 &   0.0817 &   39.296 & 2100 &  34.5 \\
   9504.76620 &   8.3426 &   0.0019 &   6.5036 &   0.0798 &   39.304 & 2100 &  52.6 \\
   9505.77836 &   8.3443 &   0.0020 &   6.4825 &   0.0727 &   39.424 & 2100 &  51.7 \\
   9506.76111 &   8.3513 &   0.0019 &   6.4833 &   0.0532 &   39.397 & 2100 &  51.5 \\
   9515.84526 &   8.3553 &   0.0020 &   6.4251 &   0.0505 &   39.651 & 2100 &  51.8 \\
   9520.79982 &   8.3479 &   0.0022 &   6.5727 &   0.0768 &   38.986 & 2100 &  47.8 \\
   9521.83149 &   8.3351 &   0.0027 &   6.5614 &   0.0698 &   39.099 & 1800 &  40.8 \\
   9528.73056 &   8.3472 &   0.0016 &   6.4964 &   0.0595 &   39.430 & 2100 &  61.4 \\
   9529.73908 &   8.3359 &   0.0015 &   6.3965 &   0.0611 &   40.033 & 2100 &  66.8 \\
   9530.74356 &   8.3309 &   0.0017 &   6.4914 &   0.0694 &   39.560 & 2100 &  57.6 \\
   9531.71638 &   8.3378 &   0.0016 &   6.3912 &   0.0595 &   40.061 & 2100 &  63.7 \\
   9543.61494 &   8.3315 &   0.0034 &   6.4788 &   0.0713 &   39.594 & 2100 &  33.3 \\
   9545.68951 &   8.3502 &   0.0027 &   6.4897 &   0.0612 &   39.545 & 2100 &  39.4 \\
   9546.72099 &   8.3539 &   0.0022 &   6.4835 &   0.0458 &   39.584 & 2100 &  47.8 \\
   9547.69656 &   8.3478 &   0.0016 &   6.5002 &   0.0518 &   39.420 & 2100 &  61.0 \\
   9548.80186 &   8.3506 &   0.0016 &   6.5131 &   0.0586 &   39.395 & 2100 &  62.6 \\
   9550.71766 &   8.3409 &   0.0039 &   6.5141 &   0.0607 &   39.461 & 1800 &  30.9 \\
   9560.66602 &   8.3452 &   0.0017 &   6.5592 &   0.0726 &   39.081 & 2100 &  59.0 \\
   9561.67646 &   8.3432 &   0.0029 &   6.5259 &   0.0777 &   39.188 & 2100 &  38.7 \\
   9563.68304 &   8.3437 &   0.0018 &   6.5071 &   0.0648 &   39.222 & 2100 &  57.8 \\
   9564.66694 &   8.3414 &   0.0021 &   6.5121 &   0.0508 &   39.162 & 2100 &  50.6 \\
   9577.74753 &   8.3495 &   0.0016 &   6.6025 &   0.0607 &   38.779 & 2100 &  62.9 \\
   9579.75033 &   8.3417 &   0.0019 &   6.5518 &   0.0774 &   39.066 & 2100 &  53.9 \\
   9581.72990 &   8.3445 &   0.0017 &   6.5317 &   0.0701 &   39.188 & 2100 &  59.1 \\
   9583.70601 &   8.3418 &   0.0019 &   6.5170 &   0.0536 &   39.369 & 2100 &  53.7 \\
   9584.59477 &   8.3422 &   0.0015 &   6.4280 &   0.0693 &   39.868 & 2100 &  67.8 \\
   9584.72285 &   8.3367 &   0.0017 &   6.5352 &   0.0527 &   39.275 & 2100 &  60.9 \\
   9585.58600 &   8.3453 &   0.0019 &   6.5261 &   0.0634 &   39.290 & 2100 &  54.7 \\
   9585.72824 &   8.3452 &   0.0018 &   6.5310 &   0.0657 &   39.298 & 2100 &  57.8 \\
   9586.72537 &   8.3349 &   0.0020 &   6.5147 &   0.0628 &   39.340 & 2100 &  52.4 \\
   9587.72124 &   8.3372 &   0.0020 &   6.5399 &   0.0621 &   39.176 & 1800 &  52.8 \\
   9588.56786 &   8.3388 &   0.0017 &   6.4888 &   0.0641 &   39.473 & 2100 &  59.8 \\
   9588.71681 &   8.3391 &   0.0020 &   6.5125 &   0.0550 &   39.362 & 1800 &  52.0 \\
   9589.71870 &   8.3407 &   0.0027 &   6.5096 &   0.0609 &   39.411 & 1808 &  41.6 \\
   9590.59269 &   8.3467 &   0.0025 &   6.5111 &   0.0596 &   39.275 & 2100 &  44.3 \\
   9591.58055 &   8.3421 &   0.0014 &   6.5012 &   0.0467 &   39.353 & 2100 &  71.9 \\
   9591.73732 &   8.3422 &   0.0030 &   6.5125 &   0.0577 &   39.195 & 2100 &  38.8 \\
   9592.59389 &   8.3501 &   0.0014 &   6.5164 &   0.0463 &   39.230 & 2100 &  69.1 \\
   9592.72732 &   8.3519 &   0.0021 &   6.5372 &   0.0396 &   38.925 & 2100 &  51.1 \\
   9606.67799 &   8.3471 &   0.0023 &   6.4959 &   0.0590 &   39.498 & 2100 &  46.8 \\
   9609.64815 &   8.3498 &   0.0015 &   6.5043 &   0.0614 &   39.329 & 2100 &  68.4 \\
   9610.65838 &   8.3519 &   0.0017 &   6.5121 &   0.0556 &   39.229 & 2100 &  62.5 \\
   9626.55046 &   8.3360 &   0.0019 &   6.5169 &   0.0653 &   39.220 & 2100 &  54.6 \\
   9627.55019 &   8.3415 &   0.0021 &   6.4967 &   0.0683 &   39.226 & 2100 &  49.5 \\
   9628.54583 &   8.3358 &   0.0017 &   6.4972 &   0.0659 &   39.329 & 2100 &  58.1 \\
   9629.56926 &   8.3376 &   0.0016 &   6.5055 &   0.0595 &   39.418 & 2100 &  59.9 \\
\end{supertabular}

\end{small}

\onecolumn

\begin{small}
\centering
\tablefirsthead{
\hline
\hline
\multicolumn{1}{c}{BJD$_\mathrm{TBD}$} &
\multicolumn{1}{c}{RV} &
\multicolumn{1}{c}{$\sigma_\mathrm{RV}$} &
\multicolumn{1}{c}{dlW} &
\multicolumn{1}{c}{$\sigma_\mathrm{dlW}$} &
\multicolumn{1}{c}{CRX} &
\multicolumn{1}{r}{$\sigma_\mathrm{CRX}$} &
\multicolumn{1}{c}{S-index} &
\multicolumn{1}{r}{$\sigma_\mathrm{S\text{-}index}$} &
\multicolumn{1}{c}{$\mathrm{H_{\alpha}}$} &
\multicolumn{1}{c}{$\mathrm{Na\,D_{1}}$} &
\multicolumn{1}{c}{$\mathrm{Na\,D_{2}}$} \\
\multicolumn{1}{c}{$-$2450000} &
\multicolumn{1}{c}{($\mathrm{km\,s^{-1}}$)} &
\multicolumn{1}{c}{($\mathrm{km\,s^{-1}}$)} &
\multicolumn{1}{c}{($\mathrm{km\,s^{-1}\,Np^{-1}}$)} &
\multicolumn{1}{c}{($\mathrm{km\,s^{-1}\,Np^{-1}}$)} &
\multicolumn{1}{c}{($\mathrm{m^2\,s^{-2}}$)} &
\multicolumn{1}{c}{($\mathrm{m^2\,s^{-2}}$)} &
\multicolumn{1}{c}{---} &
\multicolumn{1}{c}{---} &
\multicolumn{1}{c}{---} &
\multicolumn{1}{c}{---} &
\multicolumn{1}{c}{---} \\
\hline
}
\tablehead{
\multicolumn{12}{c}
{{\bfseries \tablename\ \thetable{}} -- Continued from previous page.}\\
\hline
\hline
\multicolumn{1}{c}{BJD$_\mathrm{TBD}$} &
\multicolumn{1}{c}{RV} &
\multicolumn{1}{c}{$\sigma_\mathrm{RV}$} &
\multicolumn{1}{c}{dlW} &
\multicolumn{1}{c}{$\sigma_\mathrm{dlW}$} &
\multicolumn{1}{c}{CRX} &
\multicolumn{1}{r}{$\sigma_\mathrm{CRX}$} &
\multicolumn{1}{c}{S-index} &
\multicolumn{1}{r}{$\sigma_\mathrm{S\text{-}index}$} &
\multicolumn{1}{c}{$\mathrm{H_{\alpha}}$} &
\multicolumn{1}{c}{$\mathrm{Na\,D_{1}}$} &
\multicolumn{1}{c}{$\mathrm{Na\,D_{2}}$} \\
\multicolumn{1}{c}{$-$2450000} &
\multicolumn{1}{c}{($\mathrm{km\,s^{-1}}$)} &
\multicolumn{1}{c}{($\mathrm{km\,s^{-1}}$)} &
\multicolumn{1}{c}{($\mathrm{km\,s^{-1}\,Np^{-1}}$)} &
\multicolumn{1}{c}{($\mathrm{km\,s^{-1}\,Np^{-1}}$)} &
\multicolumn{1}{c}{($\mathrm{m^2\,s^{-2}}$)} &
\multicolumn{1}{c}{($\mathrm{m^2\,s^{-2}}$)} &
\multicolumn{1}{c}{---} &
\multicolumn{1}{c}{---} &
\multicolumn{1}{c}{---} &
\multicolumn{1}{c}{---} &
\multicolumn{1}{c}{---} \\
\hline
}
\tabletail{
\hline
\multicolumn{12}{c}{{Continued on next page}}\\
\hline
}
\tablelasttail{
\hline
}
\tablecaption{Relative radial velocities and spectral activity indicators measured from the HARPS spectra with {\tt SERVAL} and {\tt TERRA}.}
\label{table-TOI-0544-3p6_harps-0108-srv-complete_output}
\begin{supertabular}{rrrrrrrrrrrr}
   8578.51514 &   0.0032 &  0.0026 &   34.208 &   3.675 &    8.059 &  23.737 &   1.287 &   0.018 &   0.432 &   1.219 &   0.968 \\
   8579.53150 &  -0.0034 &  0.0021 &   32.141 &   2.956 &    3.908 &  19.040 &   1.202 &   0.016 &   0.473 &   1.220 &   0.965 \\
   9205.63053 &   0.0194 &  0.0017 &    1.408 &   2.658 &    1.540 &  14.505 &   1.172 &   0.013 &   0.480 &   1.244 &   1.004 \\
   9206.63247 &   0.0304 &  0.0014 &    7.962 &   1.862 &   -1.187 &  12.339 &   1.126 &   0.010 &   0.501 &   1.239 &   0.988 \\
   9207.57568 &   0.0215 &  0.0019 &   14.317 &   2.154 &   11.742 &  16.447 &   1.150 &   0.012 &   0.477 &   1.241 &   0.978 \\
   9207.66046 &   0.0227 &  0.0013 &   18.549 &   1.854 &   -6.042 &  11.509 &   1.157 &   0.011 &   0.477 &   1.241 &   0.979 \\
   9214.71815 &  -0.0021 &  0.0014 &   16.143 &   1.665 &    7.481 &  11.826 &   1.098 &   0.011 &   0.496 &   1.240 &   0.990 \\
   9215.64475 &   0.0003 &  0.0012 &    7.563 &   1.671 &    3.814 &  10.529 &   1.094 &   0.009 &   0.505 &   1.243 &   0.990 \\
   9217.68640 &   0.0025 &  0.0013 &   -6.021 &   1.627 &   -3.028 &  11.113 &   1.119 &   0.011 &   0.497 &   1.243 &   0.996 \\
   9218.68225 &   0.0052 &  0.0016 &   -0.511 &   1.910 &  -16.867 &  13.649 &   1.117 &   0.010 &   0.505 &   1.248 &   0.995 \\
   9219.65704 &   0.0017 &  0.0011 &   -2.072 &   1.314 &   -7.816 &   9.266 &   1.178 &   0.010 &   0.511 &   1.245 &   0.994 \\
   9221.60896 &   0.0009 &  0.0021 &   -9.785 &   1.701 &  -11.633 &  18.359 &   1.061 &   0.010 &   0.524 &   1.247 &   0.998 \\
   9222.61015 &  -0.0022 &  0.0012 &   -9.937 &   1.485 &    1.309 &  10.524 &   1.078 &   0.009 &   0.513 &   1.255 &   0.997 \\
   9223.56895 &   0.0063 &  0.0016 &  -13.314 &   1.676 &    5.411 &  13.658 &   1.067 &   0.009 &   0.513 &   1.247 &   0.993 \\
   9224.66129 &   0.0076 &  0.0015 &   -8.005 &   1.551 &    8.850 &  12.633 &   1.125 &   0.011 &   0.496 &   1.248 &   1.001 \\
   9226.65544 &   0.0177 &  0.0018 &   15.761 &   1.684 &   -6.285 &  15.531 &   1.202 &   0.012 &   0.462 &   1.245 &   0.995 \\
   9227.66789 &   0.0076 &  0.0016 &   16.473 &   2.468 &   13.923 &  14.013 &   1.158 &   0.013 &   0.485 &   1.232 &   0.989 \\
   9228.64010 &  -0.0037 &  0.0018 &   21.783 &   2.357 &   14.554 &  15.479 &   1.222 &   0.014 &   0.460 &   1.232 &   0.999 \\
   9230.63045 &  -0.0030 &  0.0015 &   11.931 &   1.842 &    1.778 &  13.448 &   1.205 &   0.010 &   0.463 &   1.237 &   0.996 \\
   9231.63611 &  -0.0016 &  0.0018 &    5.839 &   2.627 &    4.030 &  15.996 &   1.227 &   0.013 &   0.437 &   1.241 &   0.983 \\
   9232.63467 &   0.0054 &  0.0014 &    6.978 &   1.499 &    2.628 &  12.142 &   1.148 &   0.009 &   0.508 &   1.246 &   0.979 \\
   9233.67770 &  -0.0010 &  0.0012 &    6.140 &   1.694 &   -5.641 &  10.425 &   1.227 &   0.011 &   0.499 &   1.248 &   0.980 \\
   9242.62501 &   0.0035 &  0.0011 &    1.181 &   1.876 &    1.219 &   9.600 &   1.110 &   0.009 &   0.473 &   1.248 &   0.988 \\
   9243.56834 &   0.0084 &  0.0016 &    2.409 &   1.752 &   24.240 &  13.777 &   1.042 &   0.009 &   0.500 &   1.249 &   0.997 \\
   9244.54494 &   0.0136 &  0.0014 &    5.752 &   1.937 &  -17.013 &  12.236 &   1.167 &   0.010 &   0.463 &   1.242 &   0.988 \\
   9244.62785 &   0.0105 &  0.0013 &    6.332 &   1.635 &  -17.780 &  11.490 &   1.130 &   0.010 &   0.481 &   1.245 &   0.993 \\
   9245.60955 &   0.0101 &  0.0022 &   17.823 &   2.499 &   -8.225 &  19.301 &   1.085 &   0.012 &   0.481 &   1.241 &   0.985 \\
   9246.63208 &   0.0176 &  0.0011 &   17.052 &   2.235 &   14.530 &   9.895 &   1.151 &   0.011 &   0.476 &   1.241 &   0.986 \\
   9247.59527 &   0.0067 &  0.0014 &   11.328 &   2.117 &    4.072 &  12.566 &   1.266 &   0.012 &   0.428 &   1.241 &   0.990 \\
   9248.61373 &   0.0059 &  0.0021 &   16.507 &   3.031 &    2.775 &  18.452 &   1.107 &   0.013 &   0.472 &   1.241 &   1.006 \\
   9249.62657 &   0.0061 &  0.0014 &    5.621 &   2.045 &   -0.665 &  12.126 &   1.225 &   0.012 &   0.434 &   1.237 &   0.989 \\
   9250.57953 &   0.0032 &  0.0020 &    2.527 &   2.567 &  -29.112 &  17.551 &   1.175 &   0.014 &   0.486 &   1.248 &   0.986 \\
   9251.59236 &   0.0169 &  0.0018 &    6.505 &   2.333 &   32.852 &  15.445 &   1.122 &   0.013 &   0.483 &   1.242 &   0.987 \\
   9256.62889 &   0.0116 &  0.0015 &   -1.067 &   1.991 &   20.526 &  13.127 &   1.126 &   0.012 &   0.510 &   1.254 &   0.989 \\
   9257.58116 &   0.0108 &  0.0012 &   -0.888 &   1.812 &   -3.710 &  10.565 &   1.080 &   0.010 &   0.514 &   1.249 &   0.997 \\
   9261.63836 &   0.0108 &  0.0020 &    1.164 &   2.689 &   -5.115 &  17.970 &   1.179 &   0.014 &   0.468 &   1.242 &   0.994 \\
   9262.59334 &   0.0086 &  0.0019 &    5.578 &   2.697 &   -1.354 &  16.608 &   1.136 &   0.013 &   0.484 &   1.245 &   0.991 \\
   9264.60899 &   0.0085 &  0.0015 &   12.321 &   1.688 &   23.788 &  13.227 &   1.141 &   0.011 &   0.475 &   1.243 &   0.989 \\
   9265.56638 &   0.0105 &  0.0014 &   16.936 &   1.789 &   13.431 &  12.501 &   1.144 &   0.010 &   0.479 &   1.240 &   0.996 \\
   9266.55658 &   0.0061 &  0.0016 &   21.902 &   2.065 &  -15.066 &  14.264 &   1.058 &   0.011 &   0.479 &   1.227 &   0.988 \\
   9267.55362 &  -0.0023 &  0.0017 &   19.330 &   2.024 &  -23.256 &  14.982 &   1.128 &   0.010 &   0.466 &   1.236 &   0.992 \\
   9269.57020 &   0.0036 &  0.0021 &   26.274 &   2.322 &  -13.285 &  18.773 &   1.109 &   0.012 &   0.469 &   1.233 &   0.996 \\
   9272.57672 &  -0.0034 &  0.0017 &   20.085 &   2.112 &  -25.857 &  14.051 &   1.023 &   0.011 &   0.481 &   1.245 &   0.988 \\
   9273.59284 &  -0.0062 &  0.0023 &   21.943 &   3.283 &   28.042 &  19.846 &   1.157 &   0.015 &   0.479 &   1.233 &   0.991 \\
   9274.58962 &  -0.0029 &  0.0018 &   14.123 &   1.905 &    4.171 &  15.381 &   1.048 &   0.012 &   0.482 &   1.246 &   0.987 \\
   9275.56486 &  -0.0019 &  0.0019 &    5.611 &   2.094 &   -9.529 &  16.267 &   1.098 &   0.012 &   0.509 &   1.248 &   0.996 \\
   9276.59246 &   0.0045 &  0.0020 &    8.596 &   2.098 &  -36.640 &  16.241 &   1.099 &   0.013 &   0.507 &   1.238 &   0.991 \\
   9277.57256 &   0.0101 &  0.0018 &    5.113 &   2.221 &    8.146 &  14.986 &   1.077 &   0.013 &   0.522 &   1.255 &   0.987 \\
   9284.57306 &   0.0012 &  0.0016 &    6.835 &   2.262 &   16.514 &  13.827 &   1.122 &   0.014 &   0.490 &   1.243 &   0.983 \\
   9287.53172 &   0.0040 &  0.0016 &    8.554 &   2.338 &  -11.277 &  13.664 &   1.219 &   0.013 &   0.451 &   1.240 &   0.980 \\
   9288.53857 &   0.0059 &  0.0019 &    7.805 &   2.324 &   23.854 &  16.430 &   1.196 &   0.014 &   0.465 &   1.237 &   0.980 \\
   9290.53613 &   0.0044 &  0.0021 &   13.845 &   2.513 &    3.938 &  18.318 &   1.137 &   0.013 &   0.471 &   1.236 &   0.982 \\
   9291.52201 &  -0.0008 &  0.0017 &   15.065 &   1.888 &  -11.754 &  14.296 &   1.106 &   0.012 &   0.463 &   1.233 &   0.984 \\
   9294.53123 &   0.0016 &  0.0028 &    5.710 &   3.835 &  -40.164 &  23.510 &   1.086 &   0.019 &   0.489 &   1.226 &   0.989 \\
   9295.53344 &   0.0084 &  0.0016 &   -3.504 &   2.134 &  -18.821 &  13.893 &   1.085 &   0.012 &   0.489 &   1.236 &   0.986 \\
   9296.51173 &   0.0115 &  0.0014 &   -5.641 &   1.553 &  -25.846 &  11.635 &   1.036 &   0.011 &   0.491 &   1.242 &   0.992 \\
   9297.52848 &   0.0097 &  0.0015 &    1.038 &   1.929 &  -10.220 &  12.634 &   1.080 &   0.012 &   0.497 &   1.241 &   0.999 \\
   9490.84461 &  -0.0025 &  0.0021 &  -10.560 &   2.627 &   31.414 &  17.347 &   1.017 &   0.015 &   0.494 &   1.239 &   1.000 \\
   9491.81716 &  -0.0053 &  0.0024 &   -9.314 &   3.564 &   -1.185 &  19.828 &   1.094 &   0.016 &   0.504 &   1.240 &   0.994 \\
   9497.84874 &   0.0223 &  0.0018 &   16.723 &   2.395 &   11.851 &  14.785 &   1.123 &   0.013 &   0.477 &   1.227 &   0.991 \\
   9501.84420 &   0.0081 &  0.0021 &   35.467 &   3.039 &   24.537 &  17.057 &   1.212 &   0.017 &   0.456 &   1.232 &   0.985 \\
   9502.76982 &  -0.0033 &  0.0018 &   25.389 &   2.206 &  -35.382 &  14.474 &   1.137 &   0.011 &   0.472 &   1.236 &   0.992 \\
   9503.79777 &   0.0049 &  0.0021 &   16.456 &   2.838 &   12.757 &  17.890 &   1.167 &   0.015 &   0.482 &   1.226 &   0.997 \\
   9504.76620 &  -0.0028 &  0.0015 &    1.115 &   2.058 &   -5.004 &  12.200 &   1.124 &   0.011 &   0.510 &   1.237 &   0.995 \\
   9505.77836 &   0.0011 &  0.0013 &   -8.146 &   2.128 &   -3.565 &  10.689 &   1.085 &   0.011 &   0.500 &   1.240 &   0.996 \\
   9506.76111 &   0.0124 &  0.0015 &   -7.221 &   2.400 &   -4.195 &  12.763 &   1.036 &   0.010 &   0.505 &   1.243 &   0.996 \\
   9515.84526 &   0.0169 &  0.0012 &    4.103 &   1.865 &   15.499 &   9.970 &   1.200 &   0.011 &   0.465 &   1.239 &   0.996 \\
   9520.79982 &   0.0049 &  0.0018 &   26.476 &   2.408 &   12.793 &  15.120 &   1.218 &   0.013 &   0.457 &   1.224 &   0.982 \\
   9521.83149 &  -0.0106 &  0.0023 &   20.910 &   2.396 &  -35.168 &  18.976 &   1.140 &   0.014 &   0.476 &   1.233 &   0.995 \\
   9528.73056 &   0.0038 &  0.0015 &  -16.502 &   1.675 &    9.057 &  12.269 &   1.063 &   0.010 &   0.510 &   1.248 &   1.006 \\
   9529.73908 &  -0.0047 &  0.0013 &  -15.958 &   1.503 &   10.040 &  10.958 &   1.087 &   0.009 &   0.500 &   1.246 &   1.000 \\
   9530.74356 &  -0.0128 &  0.0017 &  -19.460 &   1.694 &   12.450 &  13.925 &   0.985 &   0.010 &   0.528 &   1.247 &   1.006 \\
   9531.71638 &  -0.0043 &  0.0014 &  -22.994 &   1.857 &    2.905 &  11.465 &   1.051 &   0.009 &   0.506 &   1.253 &   1.009 \\
   9543.61494 &  -0.0062 &  0.0024 &  -11.499 &   2.915 &   60.551 &  18.773 &   1.194 &   0.017 &   0.494 &   1.242 &   0.995 \\
   9545.68951 &   0.0055 &  0.0018 &  -20.699 &   2.894 &   -0.357 &  14.722 &   1.038 &   0.013 &   0.507 &   1.233 &   0.997 \\
   9546.72099 &   0.0115 &  0.0021 &  -15.993 &   2.055 &   -3.592 &  17.166 &   1.119 &   0.012 &   0.501 &   1.249 &   0.999 \\
   9547.69656 &   0.0048 &  0.0013 &  -12.159 &   1.600 &   18.692 &  10.477 &   1.056 &   0.010 &   0.513 &   1.254 &   1.005 \\
   9548.80186 &   0.0075 &  0.0012 &  -13.524 &   1.677 &   -3.601 &  10.365 &   1.074 &   0.010 &   0.511 &   1.247 &   1.002 \\
   9550.71766 &   0.0012 &  0.0022 &  -15.334 &   3.639 &   44.748 &  17.984 &   0.987 &   0.019 &   0.503 &   1.250 &   1.008 \\
   9560.66602 &   0.0024 &  0.0015 &   15.928 &   1.898 &   -9.347 &  12.644 &   1.198 &   0.010 &   0.461 &   1.232 &   0.990 \\
   9561.67646 &  -0.0012 &  0.0020 &   10.411 &   2.569 &   24.920 &  16.595 &   1.247 &   0.015 &   0.454 &   1.239 &   0.983 \\
   9563.68304 &  -0.0004 &  0.0017 &   -1.442 &   1.830 &    7.718 &  14.258 &   1.116 &   0.010 &   0.503 &   1.243 &   0.995 \\
   9564.66694 &   0.0001 &  0.0015 &   -4.118 &   2.391 &  -14.148 &  12.000 &   1.086 &   0.011 &   0.498 &   1.245 &   1.004 \\
   9577.74753 &   0.0035 &  0.0012 &   27.385 &   1.986 &    8.929 &   9.896 &   1.255 &   0.012 &   0.452 &   1.231 &   0.982 \\
   9579.75033 &  -0.0033 &  0.0016 &   11.216 &   1.902 &   13.664 &  13.375 &   1.227 &   0.012 &   0.476 &   1.238 &   0.993 \\
   9581.72990 &   0.0017 &  0.0015 &   -5.253 &   1.439 &   -4.690 &  12.474 &   1.125 &   0.011 &   0.514 &   1.246 &   0.989 \\
   9583.70601 &  -0.0013 &  0.0016 &   -6.852 &   1.771 &    5.427 &  13.540 &   1.108 &   0.011 &   0.500 &   1.247 &   0.996 \\
   9584.59477 &   0.0000 &  0.0010 &   -7.261 &   1.347 &   -3.137 &   8.676 &   1.123 &   0.009 &   0.498 &   1.246 &   0.997 \\
   9584.72285 &  -0.0048 &  0.0014 &   -8.671 &   1.472 &   -3.113 &  11.580 &   1.119 &   0.010 &   0.503 &   1.245 &   0.993 \\
   9585.58600 &   0.0011 &  0.0018 &   -8.802 &   1.801 &  -27.674 &  14.346 &   1.117 &   0.011 &   0.509 &   1.242 &   0.996 \\
   9585.72824 &  -0.0002 &  0.0014 &   -8.467 &   1.504 &   20.429 &  11.416 &   1.060 &   0.011 &   0.519 &   1.247 &   0.991 \\
   9586.72537 &  -0.0066 &  0.0014 &   -9.467 &   2.168 &  -16.669 &  11.553 &   1.087 &   0.012 &   0.502 &   1.243 &   0.993 \\
   9587.72124 &  -0.0044 &  0.0019 &  -15.800 &   1.706 &   19.157 &  15.762 &   1.036 &   0.013 &   0.532 &   1.252 &   0.997 \\
   9588.56786 &  -0.0024 &  0.0012 &  -15.175 &   1.775 &  -10.650 &   9.543 &   1.043 &   0.009 &   0.526 &   1.251 &   0.995 \\
   9588.71681 &  -0.0035 &  0.0013 &  -18.477 &   1.962 &   -9.275 &  10.962 &   1.024 &   0.012 &   0.546 &   1.254 &   0.995 \\
   9589.71870 &  -0.0024 &  0.0020 &  -22.861 &   3.204 &    9.360 &  17.309 &   0.963 &   0.015 &   0.556 &   1.259 &   0.980 \\
   9590.59269 &   0.0045 &  0.0018 &  -16.850 &   2.585 &  -20.246 &  15.141 &   1.009 &   0.014 &   0.553 &   1.254 &   0.980 \\
   9591.58055 &  -0.0001 &  0.0012 &  -17.048 &   1.674 &  -10.154 &   9.889 &   1.052 &   0.009 &   0.545 &   1.257 &   0.968 \\
   9591.73732 &   0.0042 &  0.0021 &  -19.114 &   3.110 &   20.262 &  17.930 &   1.027 &   0.017 &   0.558 &   1.259 &   0.957 \\
   9592.59389 &   0.0097 &  0.0014 &   -9.301 &   1.424 &   12.005 &  11.847 &   1.055 &   0.009 &   0.540 &   1.255 &   0.971 \\
   9592.72732 &   0.0102 &  0.0017 &   -1.401 &   2.003 &   13.551 &  14.051 &   1.091 &   0.013 &   0.554 &   1.259 &   0.962 \\
   9606.67799 &   0.0067 &  0.0021 &  -20.079 &   2.204 &   -9.169 &  17.897 &   1.074 &   0.013 &   0.514 &   1.250 &   0.994 \\
   9609.64815 &   0.0041 &  0.0011 &  -21.853 &   1.758 &    4.793 &   8.854 &   1.100 &   0.009 &   0.489 &   1.251 &   1.005 \\
   9610.65838 &   0.0077 &  0.0015 &  -19.701 &   2.086 &    1.889 &  12.824 &   1.070 &   0.010 &   0.521 &   1.257 &   0.994 \\
   9626.55046 &  -0.0051 &  0.0014 &   -4.369 &   1.641 &    4.147 &  11.515 &   1.048 &   0.011 &   0.515 &   1.248 &   1.001 \\
   9627.55019 &  -0.0022 &  0.0017 &   -1.363 &   2.395 &   31.082 &  13.849 &   1.091 &   0.011 &   0.495 &   1.249 &   0.999 \\
   9628.54583 &  -0.0063 &  0.0014 &  -12.073 &   1.657 &  -24.609 &  11.074 &   1.065 &   0.010 &   0.518 &   1.246 &   1.005 \\
   9629.56926 &  -0.0043 &  0.0015 &  -17.272 &   1.985 &  -17.906 &  12.741 &   1.092 &   0.010 &   0.512 &   1.253 &   1.002 \\
\end{supertabular}
\end{small}

\onecolumn

\begin{small}
\centering
\tablefirsthead{
\hline
\hline
\multicolumn{1}{c}{BJD$_\mathrm{TBD}$} &
\multicolumn{1}{c}{RV} &
\multicolumn{1}{c}{$\sigma_\mathrm{RV}$} &
\multicolumn{1}{c}{FWHM} &
\multicolumn{1}{c}{BIS} &
\multicolumn{1}{c}{Contrast} &
\multicolumn{1}{c}{$\mathrm{T_{exp}}$} & 
\multicolumn{1}{c}{SNR} \\
\multicolumn{1}{c}{$-$2450000} &
\multicolumn{1}{c}{($\mathrm{km\,s^{-1}}$)} &
\multicolumn{1}{c}{($\mathrm{km\,s^{-1}}$)} &
\multicolumn{1}{c}{($\mathrm{km\,s^{-1}}$)} &
\multicolumn{1}{c}{($\mathrm{km\,s^{-1}}$)} &
\multicolumn{1}{c}{(\%)} &
\multicolumn{1}{c}{(s)} &
\multicolumn{1}{c}{@550nm} \\
\hline
}
\tablehead{
\multicolumn{8}{c}
{{\bfseries \tablename\ \thetable{}} -- continued from previous page.}\\
\hline
\hline
\multicolumn{1}{c}{BJD$_\mathrm{TBD}$} &
\multicolumn{1}{c}{RV} &
\multicolumn{1}{c}{$\sigma_\mathrm{RV}$} &
\multicolumn{1}{c}{FWHM} &
\multicolumn{1}{c}{BIS} &
\multicolumn{1}{c}{Contrast} &
\multicolumn{1}{c}{$\mathrm{T_{exp}}$} & 
\multicolumn{1}{c}{SNR} \\
\multicolumn{1}{c}{$-$2450000} &
\multicolumn{1}{c}{($\mathrm{km\,s^{-1}}$)} &
\multicolumn{1}{c}{($\mathrm{km\,s^{-1}}$)} &
\multicolumn{1}{c}{($\mathrm{km\,s^{-1}}$)} &
\multicolumn{1}{c}{($\mathrm{km\,s^{-1}}$)} &
\multicolumn{1}{c}{(\%)} &
\multicolumn{1}{c}{(s)} &
\multicolumn{1}{c}{@550nm} \\
\hline
}
\tabletail{
\hline
\multicolumn{8}{c}{{Continued on next page}}\\
\hline
}
\tablelasttail{
\hline
}
\tablecaption{Absolute radial velocities and spectral activity indicators measured from the HARPS-N spectra with the {\tt DRS}.
\label{table-TOI-0544-tng_harpn-0014-drs-complete_output}}
\begin{supertabular}{cccccccc}
   8578.35923 &   8.3519 &   0.0019 &   6.5098 &   0.0696 &   39.407 & 2400 &  51.6 \\
   8579.35283 &   8.3376 &   0.0048 &   6.4946 &   0.0508 &   39.360 & 1230 &  25.1 \\
   8732.71766 &   8.3678 &   0.0052 &   6.4912 &   0.0487 &   39.276 & 1800 &  23.1 \\
   8732.74144 &   8.3591 &   0.0064 &   6.5139 &   0.0486 &   39.264 & 1800 &  19.8 \\
   8752.68369 &   8.3692 &   0.0017 &   6.4828 &   0.0564 &   39.660 & 1800 &  57.7 \\
   8752.70581 &   8.3665 &   0.0014 &   6.4835 &   0.0600 &   39.669 & 1800 &  67.4 \\
   8753.70744 &   8.3513 &   0.0021 &   6.4844 &   0.0657 &   39.616 & 1800 &  47.3 \\
   8754.71549 &   8.3551 &   0.0017 &   6.4857 &   0.0632 &   39.663 & 1800 &  55.7 \\
   9204.51620 &   8.3541 &   0.0023 &   6.4273 &   0.0419 &   39.997 & 1800 &  42.4 \\
   9204.65656 &   8.3560 &   0.0018 &   6.4217 &   0.0470 &   40.037 & 1800 &  52.0 \\
   9205.54599 &   8.3619 &   0.0015 &   6.4309 &   0.0386 &   40.037 & 1500 &  60.1 \\
   9205.63930 &   8.3590 &   0.0014 &   6.4366 &   0.0510 &   40.001 & 1500 &  65.9 \\
   9206.47971 &   8.3674 &   0.0018 &   6.4567 &   0.0291 &   39.875 & 1800 &  52.3 \\
   9206.65731 &   8.3705 &   0.0019 &   6.4491 &   0.0362 &   39.874 & 2400 &  51.4 \\
\end{supertabular}

\end{small}

\onecolumn

\begin{small}
\centering
\tablefirsthead{
\hline
\hline
\multicolumn{1}{c}{BJD$_\mathrm{TBD}$} &
\multicolumn{1}{c}{RV} &
\multicolumn{1}{c}{$\sigma_\mathrm{RV}$} &
\multicolumn{1}{c}{dlW} &
\multicolumn{1}{c}{$\sigma_\mathrm{dlW}$} &
\multicolumn{1}{c}{CRX} &
\multicolumn{1}{r}{$\sigma_\mathrm{CRX}$} &
\multicolumn{1}{c}{S-index} &
\multicolumn{1}{r}{$\sigma_\mathrm{S\text{-}index}$} &
\multicolumn{1}{c}{$\mathrm{H_{\alpha}}$} &
\multicolumn{1}{c}{$\mathrm{Na\,D_{1}}$} &
\multicolumn{1}{c}{$\mathrm{Na\,D_{2}}$} \\
\multicolumn{1}{c}{$-$2450000} &
\multicolumn{1}{c}{($\mathrm{km\,s^{-1}}$)} &
\multicolumn{1}{c}{($\mathrm{km\,s^{-1}}$)} &
\multicolumn{1}{c}{($\mathrm{km\,s^{-1}\,Np^{-1}}$)} &
\multicolumn{1}{c}{($\mathrm{km\,s^{-1}\,Np^{-1}}$)} &
\multicolumn{1}{c}{($\mathrm{m^2\,s^{-2}}$)} &
\multicolumn{1}{c}{($\mathrm{m^2\,s^{-2}}$)} &
\multicolumn{1}{c}{---} &
\multicolumn{1}{c}{---} &
\multicolumn{1}{c}{---} &
\multicolumn{1}{c}{---} &
\multicolumn{1}{c}{---} \\
\hline
}
\tablehead{
\multicolumn{12}{c}
{{\bfseries \tablename\ \thetable{}} -- Continued from previous page.}\\
\hline
\hline
\multicolumn{1}{c}{BJD$_\mathrm{TBD}$} &
\multicolumn{1}{c}{RV} &
\multicolumn{1}{c}{$\sigma_\mathrm{RV}$} &
\multicolumn{1}{c}{dlW} &
\multicolumn{1}{c}{$\sigma_\mathrm{dlW}$} &
\multicolumn{1}{c}{CRX} &
\multicolumn{1}{r}{$\sigma_\mathrm{CRX}$} &
\multicolumn{1}{c}{S-index} &
\multicolumn{1}{r}{$\sigma_\mathrm{S\text{-}index}$} &
\multicolumn{1}{c}{$\mathrm{H_{\alpha}}$} &
\multicolumn{1}{c}{$\mathrm{Na\,D_{1}}$} &
\multicolumn{1}{c}{$\mathrm{Na\,D_{2}}$} \\
\multicolumn{1}{c}{$-$2450000} &
\multicolumn{1}{c}{($\mathrm{km\,s^{-1}}$)} &
\multicolumn{1}{c}{($\mathrm{km\,s^{-1}}$)} &
\multicolumn{1}{c}{($\mathrm{km\,s^{-1}\,Np^{-1}}$)} &
\multicolumn{1}{c}{($\mathrm{km\,s^{-1}\,Np^{-1}}$)} &
\multicolumn{1}{c}{($\mathrm{m^2\,s^{-2}}$)} &
\multicolumn{1}{c}{($\mathrm{m^2\,s^{-2}}$)} &
\multicolumn{1}{c}{---} &
\multicolumn{1}{c}{---} &
\multicolumn{1}{c}{---} &
\multicolumn{1}{c}{---} &
\multicolumn{1}{c}{---} \\
\hline
}
\tabletail{
\hline
\multicolumn{12}{c}{{Continued on next page}}\\
\hline
}
\tablelasttail{
\hline
}
\tablecaption{Relative radial velocities and spectral activity indicators measured from the HARPS-N spectra with {\tt SERVAL} and {\tt TERRA}.}
\label{table-TOI-0544-tng_harpn-0014-srv-complete_output}
\begin{supertabular}{rrrrrrrrrrrr}
   8578.35923 &  -0.0153 &  0.0012 &  -10.331 &   3.298 &  -14.213 &  10.557 &   1.289 &   0.010 &   0.382 &  1.2058 &  0.9705 \\
   8579.35283 &  -0.0189 &  0.0023 &   -2.017 &   4.584 &   -4.860 &  20.180 &   1.099 &   0.018 &   0.407 &  1.2022 &  0.9588 \\
   8732.71766 &   0.0062 &  0.0025 &   -0.417 &   5.610 &  -35.894 &  20.131 &   1.477 &   0.022 &   0.368 &  1.1750 &  0.9311 \\
   8732.74144 &  -0.0029 &  0.0033 &   -6.190 &   5.519 &  -23.484 &  28.098 &   1.482 &   0.025 &   0.374 &  1.1705 &  0.9313 \\
   8752.68369 &   0.0004 &  0.0010 &  -24.896 &   2.236 &    1.003 &   8.720 &   1.216 &   0.009 &   0.416 &  1.2200 &  0.9864 \\
   8752.70581 &   0.0001 &  0.0009 &  -27.621 &   2.072 &    9.871 &   7.201 &   1.208 &   0.007 &   0.412 &  1.2189 &  0.9840 \\
   8753.70744 &  -0.0153 &  0.0012 &  -22.886 &   2.317 &   -4.919 &  10.278 &   1.217 &   0.010 &   0.411 &  1.2071 &  0.9789 \\
   8754.71549 &  -0.0119 &  0.0011 &  -31.619 &   2.284 &  -11.948 &   9.137 &   1.264 &   0.009 &   0.396 &  1.2214 &  0.9830 \\
   9204.51620 &  -0.0115 &  0.0016 &  -57.931 &   2.722 &   30.076 &  13.728 &   1.144 &   0.011 &   0.461 &  1.2207 &  0.9819 \\
   9204.65656 &  -0.0069 &  0.0011 &  -62.660 &   2.495 &   -8.303 &   9.456 &   1.153 &   0.009 &   0.451 &  1.2261 &  0.9966 \\
   9205.54599 &  -0.0019 &  0.0011 &  -58.627 &   2.216 &   13.321 &   9.308 &   1.159 &   0.007 &   0.434 &  1.2321 &  0.9950 \\
   9205.63930 &  -0.0044 &  0.0010 &  -58.836 &   1.951 &    4.622 &   8.335 &   1.167 &   0.007 &   0.442 &  1.2318 &  0.9930 \\
   9206.47971 &   0.0017 &  0.0012 &  -47.854 &   2.351 &   -2.439 &  11.255 &   1.144 &   0.008 &   0.451 &  1.2231 &  0.9867 \\
   9206.65731 &   0.0022 &  0.0012 &  -44.826 &   2.328 &  -11.085 &   9.886 &   1.133 &   0.009 &   0.446 &  1.2166 &  0.9821 \\
\end{supertabular}

\end{small}

\newpage

\begin{table}
\caption{Comparison of all RV fitting models used, all results are within 1-sigma.}
\label{tab:resultscomp}
\begin{tabular}{@{}lllllll@{}l}
\toprule
\toprule
Toolkit                   & GP Kernel                &Activity Indicator& Number of planets & Eccentricity           & K$_\mathrm{b}$ [m/s] &M$_\mathrm{b}$ [\mearth]\\ \midrule
RadVel                    & No GP                    &-& 2                 & Circular               & $ 2.77^{+0.81}_{-0.82}$  &$3.69^{+1.19}_{1.14}$\\
RadVel                    & Celerite Quasi-Periodic  &-& 1                 & Circular               & 2.61  $\pm$  0.49         &$3.48^{+0.73}_{0.70}$\\
RadVel                    & Celerite Quasi-Periodic  &-& 2                 & Circular               & $ 2.59^{+0.49}_{-0.48}$  &$3.45^{+0.73}_{0.71}$\\
RadVel                    & Celerite Quasi-Periodic  &-& 2                 & Both eccentric         & $ 3.22^{+1.9}_{-0.69}$    &$4.29^{+2.66}_{0.98}$\\
RadVel                    & Celerite Quasi-Periodic  &-& 3                 & Circular               & 2.60 $\pm$  0.50         &$3.46^{+0.74}_{0.72}$\\
RadVel HARPS only         & Celerite Quasi-Periodic  &-& 2                 & Circular               & 2.83 $\pm$  0.53         &$3.77^{+0.79}_{0.76}$\\
RadVel                    & Square Exponential       &-& 2                 & Circular               & 2.70 $\pm$  0.40         &$3.59^{+0.61}_{0.59}$\\
RadVel                    & Square Exponential       &-& 2                 & Outer planet eccentric & $ 2.69^{+0.37}_{-0.36}$  &$3.58^{+0.57}_{0.55}$\\
RadVel                    & Square Exponential       &-& 3                 & Circular               & $ 2.67^{+0.39}_{-0.40}$  &$3.55^{+0.61}_{0.59}$\\
Pyaneti multi-dimensional & Quasi-Periodic           &S-index& 2           & Circular               & 2.07 $\pm$  0.39        & 2.75 $\pm$ 0.52\\
Pyaneti multi-dimensional & Quasi-Periodic           &S-index& 2           & Both eccentric         & $ 2.66^{+0.44}_{-0.41}$ & 3.28 $\pm$ 0.48\\
Pyaneti multi-dimensional & Quasi-Periodic           &S-index& 2           & Outer planet eccentric & 2.17 $\pm$  0.36        & 2.89 $\pm$ 0.48\\
 Pyaneti multi-dimensional & Quasi-Periodic          & Contrast& 2          & Both eccentric        &$ 2.47^{+0.41}_{-0.40}$ & 3.13 $\pm$ 0.49\\
 Pyaneti multi-dimensional & Quasi-Periodic          & FWHM& 2              & Both eccentric        & $2.48^{+0.38}_{-0.37}$ & 3.16 $\pm$ 0.47\\
 Pyaneti multi-dimensional & Quasi-Periodic          & S-index \& FWHM & 2  & Both eccentric        &$ 2.56^{+0.43}_{-0.39}$ & $3.20^{+0.48}_{-0.47}$ \\
 Pyaneti multi-dimensional & Quasi-Periodic          & S-index \& Contrast& 2& Both eccentric       & $ 2.59^{+0.42}_{-0.40}$ & $3.22^{+0.48}_{-0.46}$\\
 Pyaneti multi-dimensional & Quasi-Periodic          & FWHM \& Contrast& 2  & Both eccentric        &$ 2.55^{+0.46}_{-0.43}$  & $3.26^{+0.57}_{-0.55}$\\
Pyaneti RV only           & Quasi-Periodic           &-& 2                  & Both eccentric        & $ 2.52^{+0.40}_{-0.38}$ & $3.16^{+0.46}_{-0.47}$\\ 
Pyaneti RV only           & Quasi-Periodic           &-& 2                  & Outer planet eccentric& 2.12 $\pm$  0.35        & $2.82^{+0.48}_{-0.47}$\\
Pyaneti Joint Model       & Quasi-Periodic           &S-index& 2            & Both eccentric         & 2.58 $\pm$ 0.38       & $3.28^{+0.47}_{-0.48}$\\\bottomrule
\end{tabular}
\end{table}

\begin{table}
    \caption{Priors used for the RV fitting for each parameter. $\mathcal{U}$[a,b] refers to uniform priors in the range a – b, $\mathcal{N}$[a,b] refers to a Gaussian prior with mean a and width b, and $\mathcal{F}$[a,b] is a fixed parameter at value a.}
    \label{tab:priors}
    \begin{tabular}{ll}
    \toprule
    \toprule
    Fitted Parameter & Prior\\ \midrule
    Planet Parameters \\\midrule
    Orbital Period, $P_\mathrm{b}$ [days]& $\mathcal{N}[1.5484 , 0.000002] $ \\
    Time of Inf. Conjunction,  $T_\mathrm{conj, b}$ [BJD$_\mathrm{TDB}$] & $\mathcal{N}[2459199.0314 , 0.0007]$  \\
    RV Amplitude,  $K_\mathrm{b}$ [km s$^{-1}$] &  $\mathcal{U}[ 0.0000 , 0.0500]$  \\
     $ew1_\mathrm{b}$, $\sqrt{e_\mathrm{b}}\sin\omega_\mathrm{b}$ & $\mathcal{F}[ 0.0000 , 0.0000]$ \\
     $ew2_\mathrm{b}$, $\sqrt{e_\mathrm{b}}\cos\omega_\mathrm{b}$ & $\mathcal{F}[ 0.0000 , 0.0000]$ \\
    Orbital Period,  $P_\mathrm{c}$ [days] & $\mathcal{U}[ 49.0000 , 52.0000]$  \\
    Time of Inf. Conjunction,  $T_\mathrm{conj, c}$ [BJD$_\mathrm{TDB}$] & $\mathcal{U}[2459205.0000 , 2459225.0000]$  \\
    RV Amplitude, $K_\mathrm{c}$ [km s$^{-1}$ ]  &  $\mathcal{U}[0.0000 , 0.0500]$   \\
    $ew1_\mathrm{c}$, $\sqrt{e_\mathrm{c}}\sin\omega_\mathrm{c}$& $\mathcal{U}[ -1.0000 , 1.0000]$ \\
    $ew2_\mathrm{c}$, $\sqrt{e_\mathrm{c}}\cos\omega_\mathrm{c}$& $\mathcal{U}[ -1.0000 , 1.0000]$ \\
    \midrule
    Other Parameters \\ 
    \midrule
    
    \midrule
    GP Hyperparameters \\ 
    \midrule
    $A_0$ &  $\mathcal{U}[ 0.0,0.05]$ \\
    $A_1$ &   $\mathcal{U}[0.0,0.1]$ \\
    $A_2$ &   $\mathcal{U}[0.,0.25]$ \\
    $A_3$ & $\equiv 0$  \\
    $\lambda_\mathrm{e}$ [days] &  $\mathcal{U}[1,500]$  \\
    $\lambda_\mathrm{p}$  &   $\mathcal{U}[0.1,3.0]$   \\
    Rotation Period,  P$_\mathrm{GP}$ [days] &  $\mathcal{U}[15.0,25.0]$  \\ \bottomrule
    \end{tabular}
\end{table}

\begin{table}
    \caption{Priors used for the transit fitting for each parameter. $\mathcal{U}$[a,b] refers to uniform priors in the range a – b, $\mathcal{N}$[a,b] refers to a Gaussian prior with mean a and width b, and $\mathcal{F}$[a,b] is a fixed parameter at value a.}
    \label{tab:transitpriors}
    \begin{tabular}{ll}
    \toprule
    \toprule
    Fitted Parameter & Prior\\ \midrule
    Planet Parameters \\\midrule
    Orbital Period, $P_\mathrm{b}$ [days]& $\mathcal{U}[1.547352,1.549352] $ \\
    Time of Mid-Transit,  $T_0$ [BJD$_\mathrm{TDB}$] & $\mathcal{U}[2459199.02185,2459199.041850]$  \\
    $R_\mathrm{b}/R_\star$ & $\mathcal{U}[0.023699,0.039530]$  \\
    Impact Parameter $b$ & $\mathcal{U}[0,1]$  \\
    \midrule
    Other Parameters \\ 
    \midrule
    $M_\star$ [$M_\odot$] & $\mathcal{N}[0.630, 0.018]$   \\
    $R_\star$ [$R_\odot$] & $\mathcal{N}[0.624, 0.013]$   \\
    $u_1$ & $\mathcal{N}[0.436,0.202]$   \\
    $u_2$ & $\mathcal{N}[0.217,0.155]$  \\
    \midrule
    Noise Model Parameters \\
    \midrule
    $\rho_\mathrm{GP}$ [days] &  $\mathcal{N}[10.0,5.0]$  \\
    $\sigma_\mathrm{GP}$ [ppt] &   $\mathcal{N}[1.0,0.5]$   \\
    $\log\sigma_\mathrm{S6}$ [ppt] &   $\mathcal{N}[-0.614,10]$   \\
    $\log\sigma_\mathrm{S32}$ [ppt] &   $\mathcal{N}[-0.684,10]$   \\
\bottomrule
    \end{tabular}
\end{table}


\bsp	
\label{lastpage}
\end{document}